\documentclass[pdflatex,sn-mathphys-num]{sn-jnl}%
\renewcommand{\orcid}[1]{}

\usepackage{graphicx}%
\usepackage{multirow}%
\usepackage{amsmath,amssymb,amsfonts}%
\usepackage{amsthm}%
\usepackage{mathrsfs}%
\usepackage[title]{appendix}%
\usepackage{xcolor}%
\usepackage{textcomp}%
\usepackage{manyfoot}%
\usepackage{booktabs}%
\usepackage{algorithm}%
\usepackage{algorithmicx}%
\usepackage{algpseudocode}%
\usepackage{listings}%
\usepackage{xeCJK}

\usepackage{subcaption} 
\usepackage[percent]{overpic}

\theoremstyle{thmstyleone}%
\theoremstyle{thmstyletwo}%

\theoremstyle{thmstylethree}%

\begin{document}

\title[Article Title]{Charge-Exchange Reactions Accompanied by a Single $\pi^+$ Production in Medium-Energy Heavy-Ion Collisions
}


\author[1,2]{\fnm{Xiao-lei} \sur{Chen}{（陈晓磊）}}
\email{xiaoleichen15@gmail.com}

\author[1,2]{\fnm{Jin-yu} \sur{Yang}{（杨谨渝）}}\email{1799512527@qq.com}
\author[1,2]{\fnm{Shi-kang} \sur{Dong}{（董世康）}}\email{dongshk2025@lzu.edu.cn}
\author[1,2]{\fnm{Bao-wei} \sur{Ding}{（丁宝卫）}}\email{dingbw@lzu.edu.cn}

\author*[1,2]{\fnm{Bi-tao} \sur{Hu}{（胡碧涛）}}\email{hubt@lzu.edu.cn}
\author*[1,2]{\fnm{Xi-yu} \sur{Qiu}{（邱玺玉）}}\email{qiuxy@lzu.edu.cn}

\affil[1]{\orgdiv{Frontiers Science Center for Rare Isotopes}, 
\orgname{Lanzhou University}, 
\orgaddress{\city{Lanzhou}, \postcode{730000}, \country{PR China}}}

\affil[2]{\orgdiv{School of Nuclear Science and Technology}, 
\orgname{Lanzhou University}, 
\orgaddress{\city{Lanzhou}, \postcode{730000}, \country{PR China}}}


\abstract{

Heavy-ion charge-exchange (CE) reactions provide a sensitive probe of isospin dynamics in nuclear collisions.
We investigate the reaction $^{12}\mathrm{C}(^{12}\mathrm{C},\,^{12}\mathrm{N}\,\pi^{+})\,^{12}\mathrm{Be}$ at 400--600~A\,MeV within the ultra-relativistic quantum molecular dynamics model combined with a phase-space coalescence approach.
This reaction represents a nontrivial CE channel accompanied by a single $\pi^+$ production in heavy-ion collisions, extending previous studies from lepton-induced to hadronic systems.
The $^{12}\mathrm{N}$ fragment is formed via nucleon and meson exchange, whereas $\pi^+$ production is primarily governed by $\Delta$ resonance excitation and decay, enabling simultaneous investigations of CE processes and $\Delta$-induced pion production within the same reaction system.
We calculate the reaction cross section and analyze the four-momentum distributions of $^{12}\mathrm{N}$ and $\pi^+$. 
Characteristic phase-space features reflect different production mechanisms and provide guidance for future experimental designs.
Additionally, this reaction may serve as a potential pathway for rare-isotope production.

}


\maketitle

\section{Introduction}\label{sec1}
Heavy-ion collisions offer powerful platforms for studying strong interactions and the properties of nuclear matter. 
By enabling the exploration of a wide range of incident energies, heavy-ion collisions probe nuclear matter under diverse conditions, ranging from sub-saturation densities to extremely high-temperature and high-density environments.
At low and medium energies, heavy-ion reactions provide important insights into nuclear reaction mechanisms, the equation of state (EoS) of nuclear matter, and the evolution of nuclear shell structure \cite{bib1, CPC1, bib2, bib3,MA, PRCzhang, PRCSu, PRCSu2, PRCzhyp}.
At higher energies, they enable the formation of hot and dense hadronic matter and may lead to the creation of a quark–gluon plasma (QGP) \cite{bib4, bib5, bib6, 2023, Chencpc}, providing insights into the state of matter in the early universe shortly after the Big Bang. 
In addition, high-energy collisions can produce various strange particles and hypernuclei \cite{bib7, bib8, PRCzyp, 2010gyg}. 
Studies of hypernuclei provide constraints on hyperon–nucleon (Y–N) and hyperon–hyperon (Y–Y) interactions and are essential for understanding neutron stars \cite{bib9, bib10, bib11, PRCzyp, PRLzhyp}.

Among the various reactions in heavy-ion physics, CE reactions play a unique role. 
These reactions are characterized by the transfer of charge between a projectile and target while conserving the mass number. 
At low energies, CE reactions are dominated by proton–neutron exchange mechanisms between nucleons \cite{bib12}. 
With increasing collision energy, charged-meson exchange processes become increasingly important and compete with nucleon exchange \cite{bib13}. 
Because protons and neutrons are two isospin states of the nucleon, CE reactions can be interpreted as isospin transitions induced by the strong interaction. 
Such processes are analogous to nuclear $\beta$ decay and thus provide an effective probe of $\beta$-like transitions \cite{bib14}. 
Consequently, CE reactions offer access to nuclear matrix elements (NMEs) relevant to weak-interaction processes, including neutrino–nucleus reactions and neutrinoless double-$\beta$ decay (0$\nu\beta\beta$) \cite{bib15, bib16}. 
Furthermore, CE studies contribute to a deeper understanding of meson-exchange mechanisms, nucleon–nucleon correlations, and spin–isospin excitations in nuclei \cite{bib17, bib18, bib19, bib20}.

Medium- and high-energy heavy-ion CE reactions also provide potential pathways for the production of hypernuclei, particularly neutron-rich and neutral hypernuclei \cite{bib21}, such as $^{2}_{\Lambda}\mathrm{n}$, $^{3}_{\Lambda}\mathrm{n}$, $^{6}_{\Lambda}\mathrm{H}$, and $^{7}_{\Lambda}\mathrm{H}$. Investigations of these systems address fundamental questions regarding the possible existence of neutral hypernuclei and the origin of the extremely short lifetimes of neutron-rich hypernuclei \cite{bib22}.
However, owing to their extremely small production cross sections and the scarcity of experimental data, particularly at high energies, these studies remain largely theoretical. Thus, further experimental verifications are required.
 
Among various reaction mechanisms, experimental studies have shown that reactions involving $\Delta^+$ excitation can produce the rare isotopes of interest, where the $\Delta^+$ excitation decays via $\pi^+$ and neutron emission~\cite{A12025}.
Although these processes are induced by leptons, the underlying mechanism is universal and equally applicable to heavy-ion collisions.

In this paper, we present a systematic theoretical investigation of the $^{12}\mathrm{C}(^{12}\mathrm{C},\,^{12}\mathrm{N}\,\pi^{+})\,^{12}\mathrm{Be}$ reaction using the ultra-relativistic quantum molecular dynamics (UrQMD) model. 
This reaction represents a nontrivial heavy-ion charge-exchange (CE) channel accompanied by a single $\pi^+$ production. 
Compared with conventional CE studies focusing on nucleon-induced or light-ion reactions, this paper extends the investigation to a heavy-ion system, thereby providing a new perspective on isospin dynamics and meson production in a many-body environment. 
A distinctive feature of this reaction is that both CE processes and $\Delta$ resonance excitation and decay are involved in the same reaction system, enabling the study of these mechanisms under identical reaction conditions.

Using the UrQMD model, we calculate the reaction cross section and perform a systematic analysis of the four-momentum distributions of the emitted $^{12}\mathrm{N}$ and $\pi^+$ particles. 
Multidimensional distributions in momentum, energy, emission angle, and rapidity exhibit characteristic phase-space features associated with different production mechanisms, providing detailed insights into the underlying reaction dynamics.

Furthermore, the processes may constitute a pathway for the production of rare isotopes in complex nuclear systems. 
Overall, this paper provides a consistent theoretical framework and quantitative predictions that can serve as a useful reference for future experimental investigations.

\section{Theoretical Framework}\label{sec2}

\subsection{
UrQMD Model
} 

The UrQMD model is a microscopic many-body theoretical framework based on transport theory. 
In this model, nucleons are described as Gaussian wave packets with finite spatial widths, the centroids of which in coordinate and momentum space evolve dynamically under the influence of the system Hamiltonian \cite{bib24, BLEICHER2022103920}. 
The Hamiltonian incorporates several interaction terms, including two- and three-body Skyrme interactions, Yukawa potentials, Coulomb interactions, Pauli blocking, symmetry energy contributions, and momentum-dependent potentials. 
Detailed parameter settings of these interactions are provided in Ref. \cite{bib24}. 
These potentials collectively govern the interactions among nucleons and determine their spatiotemporal evolution through the Hamiltonian equations of motion:
\begin{equation}
\frac{d\vec{r}_i}{dt} = \frac{\partial H}{\partial \vec{p}_i},
\end{equation}
\begin{equation}
\frac{d\vec{p}_i}{dt} = -\frac{\partial H}{\partial \vec{r}_i},
\end{equation}
where $\vec{r}_i$ and $\vec{p}_i$ denote the position and momentum of the $i$th nucleon, respectively. 
In addition, the UrQMD model incorporates stochastic processes such as particle scattering and decay, which are treated using experimentally measured cross sections and resonance parameters. 
Thus, the UrQMD model provides a comprehensive description of various physical mechanisms in nuclear reactions over a broad energy range, including elastic and inelastic scattering, resonance excitation and decay, and meson production. 
Consequently, it can reproduce experimentally observed particle production probabilities, rapidity distributions, and angular distributions with reasonable accuracy \cite{bib24, BLEICHER2022103920, bib28}.

To date, the UrQMD model has been extensively applied to simulate nuclear reactions over a wide energy range, from SIS to RHIC. Its applications include nucleon–nucleus collisions \cite{bib29}, nucleus–nucleus collisions \cite{bib30, bib31, PRCLi, PRCWang}, and strange particle production \cite{bib29, bib32}.
Previous studies have shown that the UrQMD model is suitable for investigating heavy-ion collisions in the medium-energy region.
In this paper, the UrQMD model is employed to simulate and analyze CE reactions accompanied by single $\pi^+$ production in medium-energy heavy-ion collisions.

The numerical simulations in this paper are performed using the UrQMD model. 
To ensure reproducibility and account for model sensitivity in the medium-energy regime, we specify the detailed settings as follows.
We simulate $^{12}\text{C}+{}^{12}\text{C}$ collisions at incident energies of $400$, $500$, and $600$~A\,MeV. 
All calculations are conducted in the laboratory reference frame. 
The impact parameter is selected as $b = 0$--$5.5~\text{fm}$, covering central to peripheral collisions.
Because the CE reactions with a single $\pi^+$ production require sufficient nuclear overlap for $\Delta$ resonance production, this range includes most of the relevant events for the $^{12}\text{C}+{}^{12}\text{C}$ system.

For the nuclear mean-field potential, a hard Skyrme-type EoS is adopted, which is suitable for describing the collective dynamics and pion production in this energy region.
The initial nuclei are prepared using a stabilization procedure consistent with the quantum molecular dynamics (QMD) mean-field, instead of a simple Woods–Saxon initialization.
This treatment ensures the structural stability and proper binding energies of the projectile and target, avoiding unphysical pre-collision oscillations and nucleon emission.

The dynamical evolution is followed up to a total propagation time of $100~\mathrm{fm}/c$.
At this time, the system reaches the low-density freeze-out stage, where fragment formation and pion production have stabilized, and the yields are saturated \cite{Li2005}.

\subsection{
Phase-space Coalescence Model
} 

The phase-space coalescence model is applied to the final-state nucleons from the UrQMD simulations to reconstruct and identify the nuclear fragments produced in the reaction.
This model has been widely used in dynamical models, such as the Boltzmann--Uehling--Uhlenbeck (BUU), QMD, and UrQMD models, to identify nuclear clusters and estimate their yields \cite{bib33, bib34, bib35}.
It has been shown to provide a reliable description of cluster formation in the medium-energy regime for nuclei with mass numbers $A = 1\text{--}14$ \cite{bib44}. 
Therefore, it is suitable for describing fragment formation in the $A = 12$ reaction system considered in this paper.

In the final state of the initial collision, emitted nucleons that are sufficiently close to each other in phase space are assumed to originate from the same primary fragment; thus, they are considered to form a cluster.
In this model, two nucleons $i$ and $j$ are considered to belong to the same cluster if the following conditions are satisfied:
\begin{equation}
|\vec{p}_i - \vec{p}_j| < p_0, \quad 
|\vec{r}_i - \vec{r}_j| < r_0,
\end{equation}
where $p_0$ and $r_0$ denote the momentum- and coordinate-space coalescence parameters, respectively. 
Typically, $p_0$ ranges from 0.25 to 0.35~GeV/$c$ and $r_0$ ranges from 3 to 4~fm \cite{bib34}. 

\begin{figure}[h]
\centering
\begin{minipage}[t]{0.48\textwidth}
    \centering
    \includegraphics[width=\textwidth]{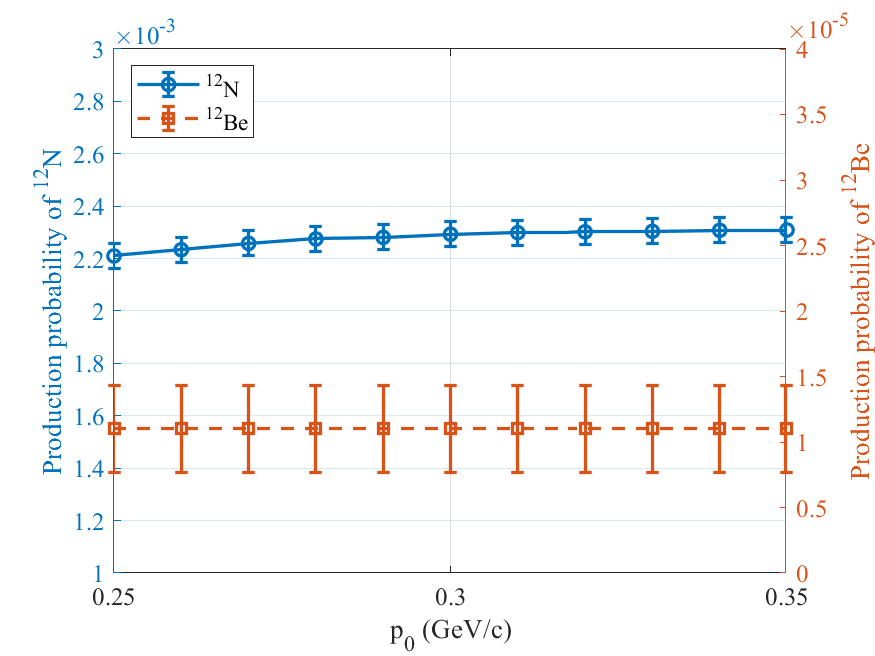}
    \caption{Influence of $p_0$ on the production probability of $^{12}\text{N}$ and $^{12}\text{Be}$ in $^{12}\text{C}+^{12}\text{C}$ collisions at $500~A\,MeV$($r_0=3.8\text{ fm})$}
    \label{figp0}
\end{minipage}
\hfill
\begin{minipage}[t]{0.48\textwidth}
    \centering
    \includegraphics[width=\textwidth]{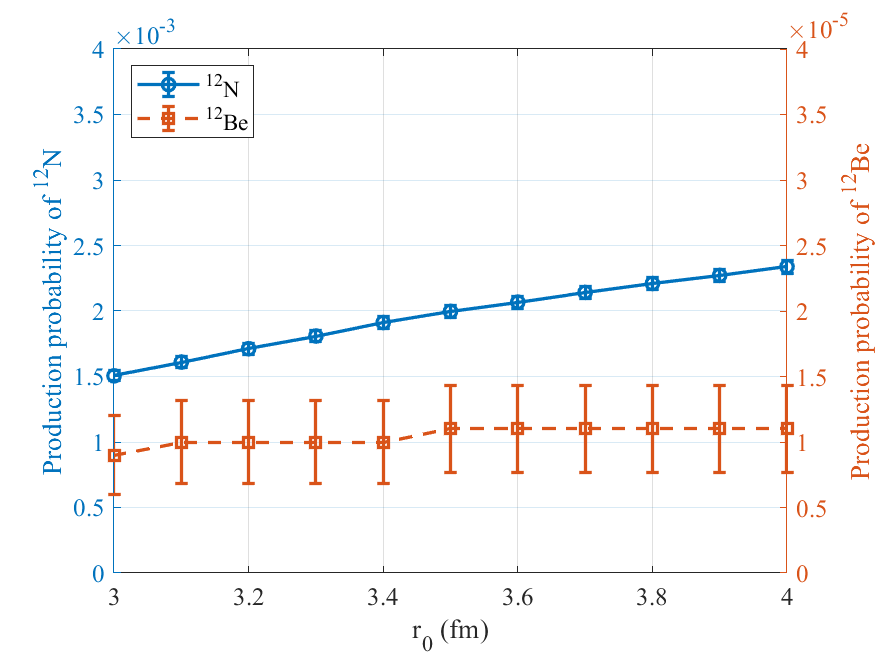}
    \caption{Influence of $r_0$ on the production probability of $^{12}\text{N}$ and $^{12}\text{Be}$ in $^{12}\text{C}+^{12}\text{C}$ collisions at $500~A\,MeV$($p_0=0.25\text{ GeV}/c)$}
    \label{figr0}
\end{minipage}
\end{figure}

To examine the sensitivity of the coalescence production probability to $p_0$ and $r_0$, we perform the simulations with $10^6$ events at an incident energy of 500~A\,MeV.
The production probabilities of $^{12}\text{N}$ and $^{12}\text{Be}$ are evaluated for different parameter sets, as shown in Figs.~\ref{figp0} and \ref{figr0}.
The error bars represent the square root of the event counts ($\sqrt{N}$), corresponding to Poisson statistical uncertainties.
The results indicate that the production probabilities are weakly dependent on $p_0$ but increase significantly with increasing $r_0$.

This behavior can be understood in terms of the underlying reaction mechanism. Because the projectile, target, and fragments share the same mass number ($A=12$), the reaction is dominated by peripheral collisions. In such collisions, nucleons undergo only small deflections and remain spatially localized, resulting in a strong dependence on the coordinate-space criterion $r_0$. In contrast, the small momentum transfer limits the variation of the relative momenta among nucleons participating in the coalescence process. As a result, varying the momentum-space criterion $p_0$ has only a minor effect on the number of nucleons satisfying the coalescence condition, leading to a weak dependence of the production probability on $p_0$.

Because the phase-space coalescence model is essentially a phenomenological tool intended for order-of-magnitude estimations and trend analysis, its parameter settings enable a reasonable range of values.
Considering the different interactions between nucleon pairs, particularly the Coulomb repulsion between protons, a smaller coalescence radius is adopted for proton–proton pairs. In this paper, we adopt the momentum-space coalescence parameter and coordinate-space coalescence radii as
\begin{equation}
p_0 = 0.25~\mathrm{GeV}/c, \quad
r_{pp}=2.8~\mathrm{fm}, \quad
r_{nn}=r_{np}=3.8~\mathrm{fm}.
\end{equation}

The momentum of a reconstructed fragment is obtained by summing the momenta of its constituent nucleons:
\begin{equation}
\vec{p} = \sum_i \vec{p}_i = (\vec{p}_x, \vec{p}_y, \vec{p}_z),
\end{equation}
and the magnitude of the total momentum is given by
\begin{equation}
p = \sqrt{\vec{p}_x^2 + \vec{p}_y^2 + \vec{p}_z^2}.
\end{equation}
The total energy is then calculated using the relativistic energy–momentum relation:
\begin{equation}
E = \sqrt{p^2 c^2 + m^2 c^4},
\end{equation}
where $m$ is the rest mass of the fragment. 
This simple and efficient method is widely used in transport-model analysis of nucleus production in medium- to high-energy heavy-ion collisions.

In this paper, we focus on the reaction channel
$^{12}\mathrm{C}(^{12}\mathrm{C},\,^{12}\mathrm{N}\,\pi^{+})\,^{12}\mathrm{Be}$
to investigate CE reactions accompanied by the production of a single $\pi^+$ meson.
This channel is particularly advantageous because $^{12}\mathrm{N}$ exists only in its ground state and has no bound excited states \cite{bib23}, thereby eliminating uncertainties associated with excitation energies.
Thus, the total energies of $^{12}\mathrm{N}$ and $\pi^+$ can be accurately determined using their rest masses, $m(^{12}\mathrm{N}) = 12.0186132~\mathrm{u} \approx 11.195~\mathrm{GeV}$ and $m(\pi^+) \approx 0.139~\mathrm{GeV}$ \cite{bib37, bib38}.
The nucleon dynamics at incident energies of 400, 500, and 600~A\,MeV are simulated using the UrQMD model, followed by the application of the coalescence model to reconstruct the final-state fragments.

\section{Results}\label{sec3}

In this paper, we employ the UrQMD model to simulate $1.44 \times 10^{9}$ events at medium energies (400, 500, and 600~A\,MeV) for the incident channel $^{12}\mathrm{C}+^{12}\mathrm{C}$.
The final-state nucleons from the collisions are subsequently processed using a phase-space coalescence model, and we calculate the probability of event production for the specific channel:
$^{12}\mathrm{C}(^{12}\mathrm{C},\,^{12}\mathrm{N}\,\pi^{+})\,^{12}\mathrm{Be}$.

The reaction cross section at each incident energy is calculated according to \cite{bib21}:
\begin{equation}
\sigma = A_{\text{reaction}} \cdot \frac{N}{N_{\text{total}}},
\end{equation}
where $\sigma$ denotes the cross section of the channel of interest, $A_{\text{reaction}}$ represents the geometric reaction area, $N_{\text{total}}$ is the total number of simulated events, and $N$ is the number of events corresponding to the selected reaction channel.
In the simulation, the impact parameter $b$ is sampled according to the geometrical prescription
\begin{equation}
d\sigma = 2\pi b\, db ,
\end{equation}
which results in a probability distribution:
\begin{equation}
\frac{dP}{db} \propto b .
\end{equation}
The sampling is performed within the range $0 \le b \le 5.5$ fm, resulting in $A_{\text{reaction}} = \pi \times (5.5 \,\mathrm{fm})^{2}$.

\begin{figure}[h]
\centering
\includegraphics[width=0.7\textwidth]{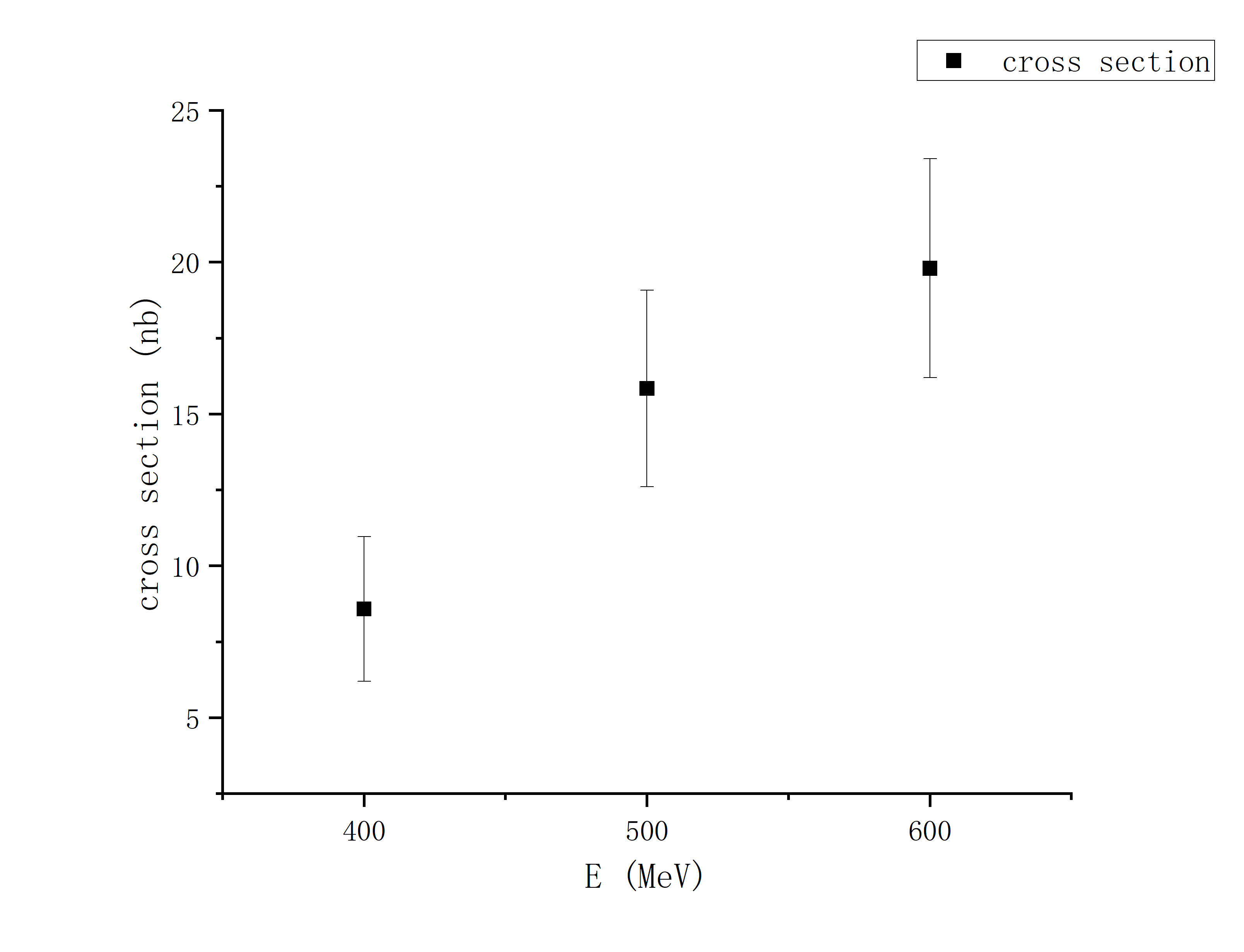} 
\caption{Cross sections of $^{12}\mathrm{C}\,(^{12}\mathrm{C},\,^{12}\mathrm{N}\,\pi^{+})\,^{12}\mathrm{Be}$ at 400, 500, and 600~A\,MeV}
\label{figcross_section}
\end{figure}

The corresponding numbers of identified events for the
$^{12}\mathrm{C}(^{12}\mathrm{C},^{12}\mathrm{N}\,\pi^{+})^{12}\mathrm{Be}$
reaction are 13, 24, and 30, resulting in cross sections of
$8.58 \pm 2.38~\mathrm{nb}$, $15.84 \pm 3.23~\mathrm{nb}$, and $19.80 \pm 3.61~\mathrm{nb}$,
respectively.
As shown in Fig.~\ref{figcross_section}, the cross section exhibits a monotonic increase with the incident energy.

We further investigate the dynamical properties of the
$^{12}\mathrm{C}(^{12}\mathrm{C},^{12}\mathrm{N}\,\pi^{+})^{12}\mathrm{Be}$ reaction. 
By analyzing the distributions of energies, momenta, and other relevant observables of the reaction products, we aim to elucidate the underlying reaction dynamics, identify characteristic phase-space features associated with different production mechanisms, and explore potential pathways for the production of rare isotopes while providing quantitative predictions to guide future experimental studies.

\subsection{
Momentum and Energy Distributions for the 
$^{12}\mathrm{C}(^{12}\mathrm{C},\,^{12}\mathrm{N}\,\pi^{+})\,^{12}\mathrm{Be}$ Reaction
} 

In this subsection, we examine the momentum and energy distributions of the emitted $^{12}\mathrm{N}$ nuclei and $\pi^+$ mesons in the laboratory frame.
This analysis provides direct insights into the kinematic characteristics of the reaction products and reflects the underlying mechanisms of energy and momentum transfer during the collision process.
To achieve a comprehensive understanding, we systematically investigate the total, longitudinal, and transverse momentum distributions, together with the corresponding energy distribution.

\begin{figure}[h]
\centering
\begin{minipage}[t]{0.48\textwidth}
    \centering
    \includegraphics[width=\textwidth]{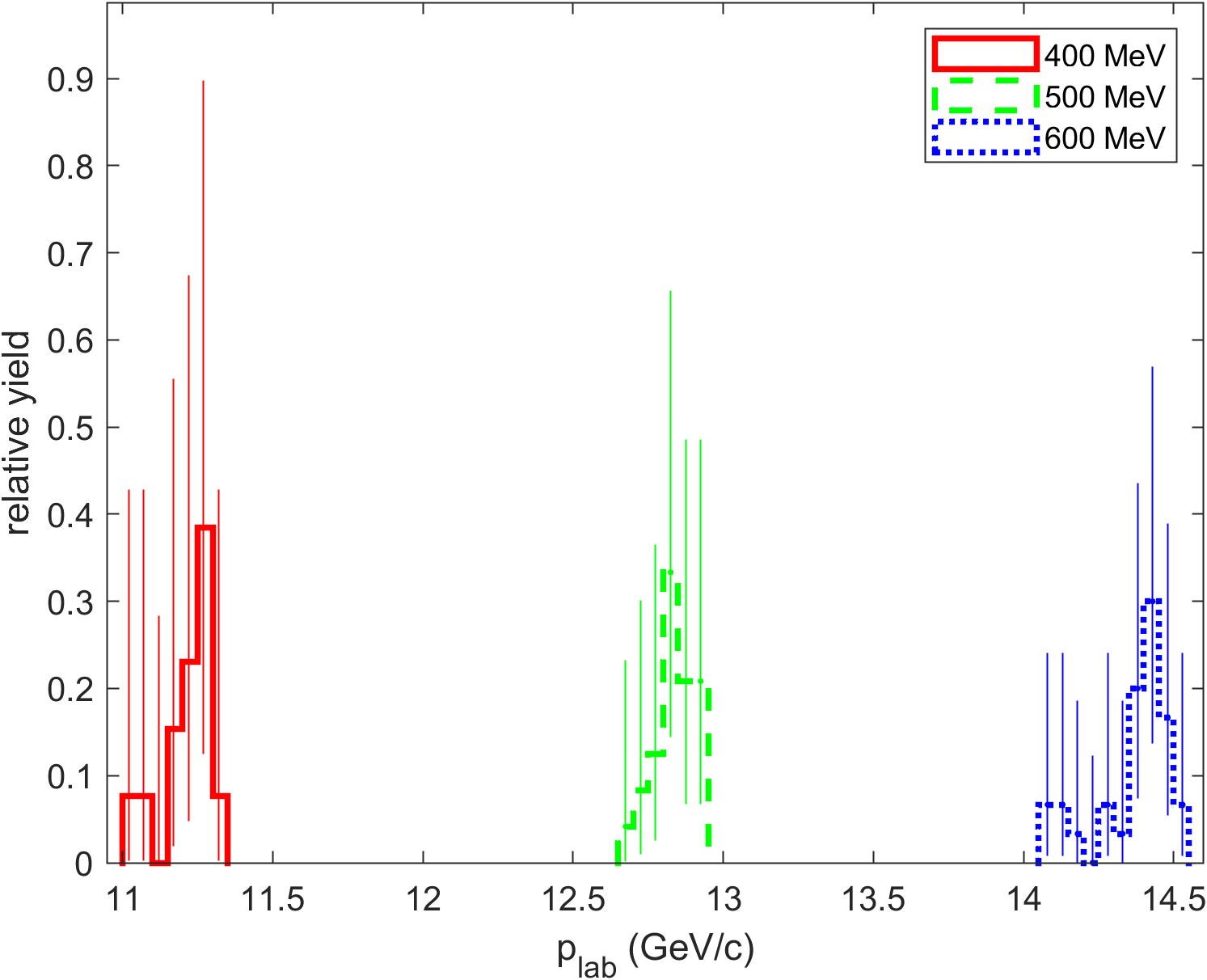}
    \caption{Normalized total momentum distributions of the emitted $^{12}\mathrm{N}$ nuclei in the laboratory frame for the reaction $^{12}\mathrm{C}\,(^{12}\mathrm{C},\,^{12}\mathrm{N}\,\pi^{+})\,^{12}\mathrm{Be}$ at incident energies of $400$, $500$, and $600~\mathrm{A\,MeV}$.}
    \label{figNptot}
\end{minipage}
\hfill
\begin{minipage}[t]{0.48\textwidth}
    \centering
    \includegraphics[width=\textwidth]{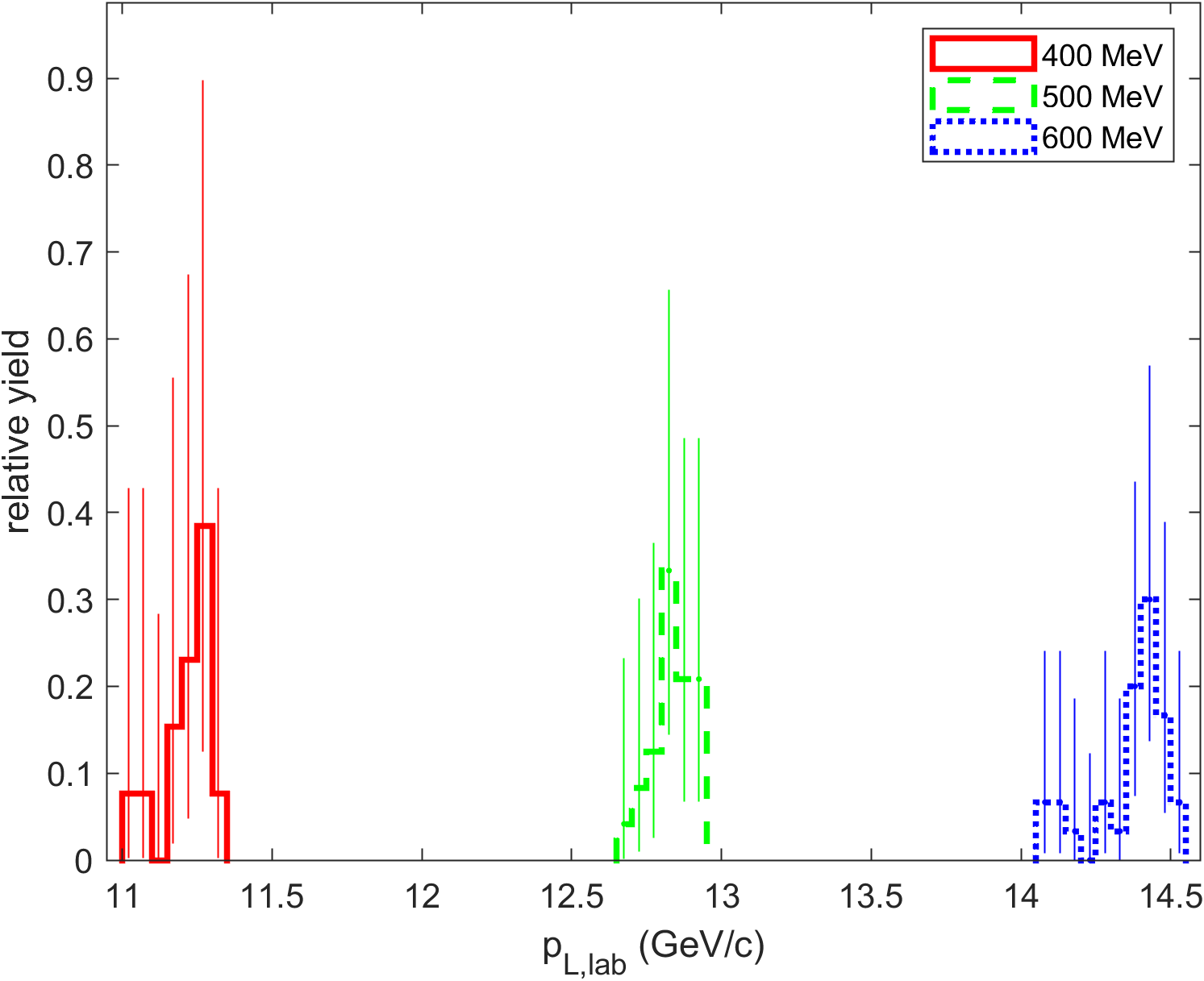}
    \caption{Normalized longitudinal momentum distributions of the emitted $^{12}\mathrm{N}$ nuclei in the laboratory frame for the reaction $^{12}\mathrm{C}\,(^{12}\mathrm{C},\,^{12}\mathrm{N}\,\pi^{+})\,^{12}\mathrm{Be}$ at incident energies of $400$, $500$, and $600~\mathrm{A\,MeV}$.}
    \label{figNplon}
\end{minipage}
\end{figure}

\begin{figure}[h]
\centering
\begin{minipage}[t]{0.48\textwidth}
    \centering
    \includegraphics[width=\textwidth]{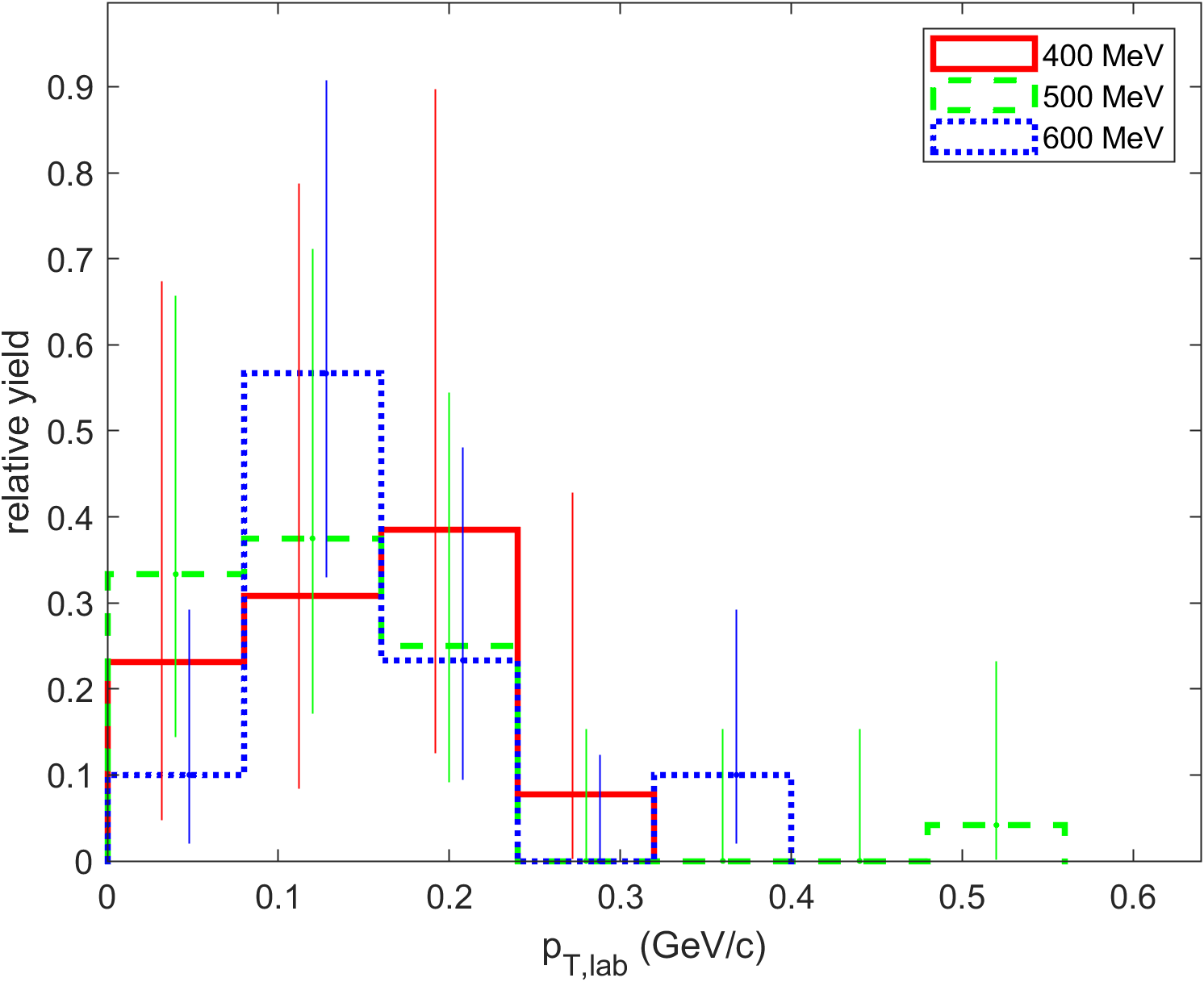}
    \caption{Normalized transverse momentum distributions of the emitted $^{12}\mathrm{N}$ nuclei in the laboratory frame for the reaction $^{12}\mathrm{C}\,(^{12}\mathrm{C},\,^{12}\mathrm{N}\,\pi^{+})\,^{12}\mathrm{Be}$ at incident energies of $400$, $500$, and $600~\mathrm{A\,MeV}$.}
    \label{figNptra}
\end{minipage}
\hfill
\begin{minipage}[t]{0.48\textwidth}
    \centering
    \includegraphics[width=\textwidth]{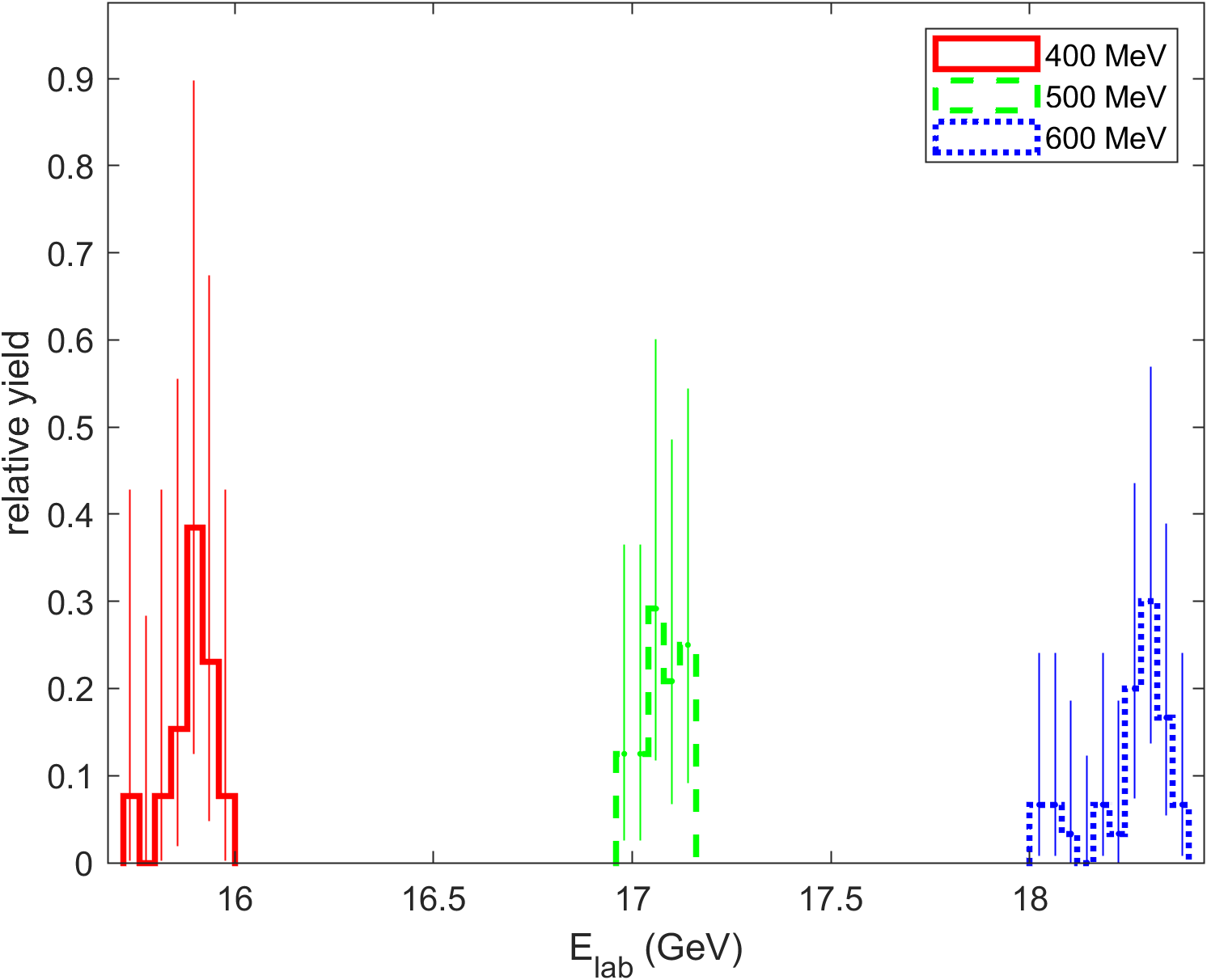}
    \caption{Normalized energy distributions of the emitted $^{12}\mathrm{N}$ nuclei in the laboratory frame for the reaction $^{12}\mathrm{C}\,(^{12}\mathrm{C},\,^{12}\mathrm{N}\,\pi^{+})\,^{12}\mathrm{Be}$ at incident energies of $400$, $500$, and $600~\mathrm{A\,MeV}$.}
    \label{figNE}
\end{minipage}
\end{figure}

The total momentum distributions of the emitted $^{12}\mathrm{N}$ nuclei in the laboratory frame are presented in Fig.~\ref{figNptot}. 
Owing to limited statistics in several datasets, statistical uncertainties are evaluated using exact Poisson confidence intervals (Garwood method) instead of the $\sqrt{N}$ approximation\cite{bib0508}. 
The confidence intervals are derived from the chi-square distribution and correspond to the $95\%$ confidence level ($\alpha = 0.05$).

With this statistical treatment, the extracted distributions exhibit clear systematic behavior. 
As the incident energy increases from 400 to 600~A\,MeV, the peak position shifts toward higher momenta, accompanied by a reduction in the peak height, whereas the overall width remains essentially unchanged.

The longitudinal momentum distributions in the laboratory frame are presented in Fig.~\ref{figNplon}. 
The normalized $p_L$ distributions exhibit characteristics similar to those of the total momentum distributions. 
In particular, as the incident energy increases, the peak position shifts toward higher momentum, accompanied by a decrease in the peak height, whereas the distribution width remains essentially unchanged.

The transverse momentum distributions are presented in Fig.~\ref{figNptra}.
In contrast to the longitudinal case, the normalized $p_T$ distributions exhibit nearly identical shapes across all three incident energies.
The distributions display pronounced maxima in the low-$p_T$ region, with the dominant contributions concentrated in $p_{T,\mathrm{lab}} \approx 0.05$--$0.25~\mathrm{GeV/c}$. 
Within the uncertainties, no significant shift of the peak position or notable change in the width is observed as the incident energy increases, indicating a weak dependence of the transverse momentum on the incident energy.

The energy distributions of the emitted $^{12}\mathrm{N}$ nuclei are presented in Fig.~\ref{figNE}. 
As the incident energy increases from 400 to 600~A\,MeV, the peak positions shift toward higher energies, whereas the overall widths remain essentially constant, and the peak heights decrease slightly.

In summary, a coherent picture emerges from the momentum and energy distributions. 
The incident energy primarily influences the longitudinal dynamics of the emitted fragments, as reflected by the systematic shift in the total and longitudinal momentum distributions, whereas the transverse momentum distributions remain largely unchanged. 
The approximately constant widths suggest that the fragmentation process largely preserves the intrinsic momentum spread of the projectile. 
These characteristics indicate that $^{12}\mathrm{N}$ production is dominated by peripheral collisions, in which the projectile undergoes limited energy dissipation, rather than by central collisions associated with substantial momentum and energy transfer.

Next, we analyze the total, longitudinal, and transverse momentum distributions, as well as the energy distribution, of the produced $\pi^+$ mesons in the laboratory frame to elucidate their production mechanism in the 
$^{12}\mathrm{C}(^{12}\mathrm{C},\,^{12}\mathrm{N}\,\pi^{+})\,^{12}\mathrm{Be}$ reaction. 
These distributions characterize the kinematic behavior of the $\pi^+$ mesons and offer insight into the underlying reaction dynamics.

\begin{figure}[h]
\centering
\begin{minipage}[t]{0.48\textwidth}
    \centering
    \includegraphics[width=\textwidth]{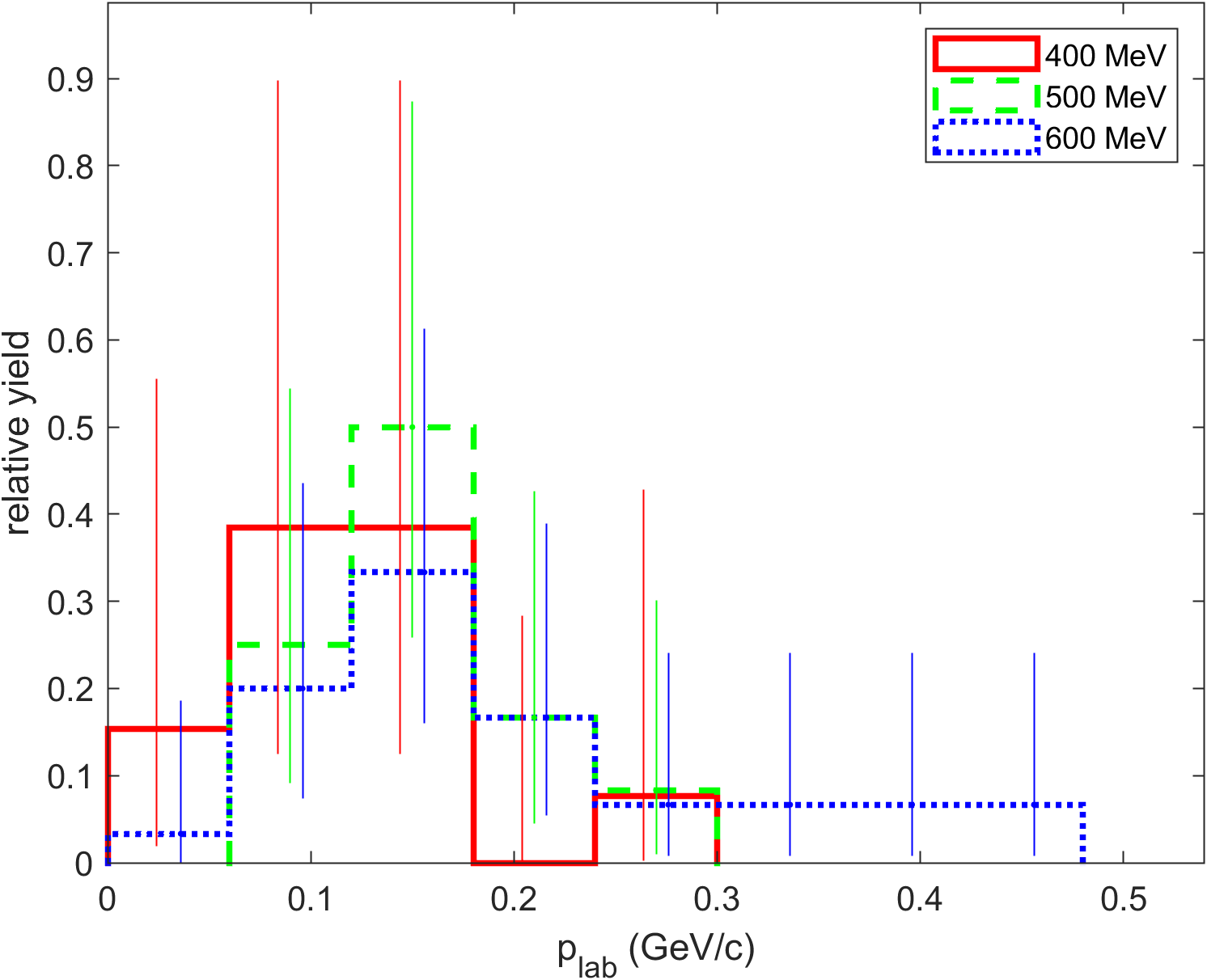}
    \caption{Normalized total momentum distributions of the produced $\pi^+$ mesons in the laboratory frame for the reaction $^{12}\mathrm{C}\,(^{12}\mathrm{C},\,^{12}\mathrm{N}\,\pi^{+})\,^{12}\mathrm{Be}$ at incident energies of $400$, $500$, and $600~\mathrm{A\,MeV}$.}
    \label{figpiptot}
\end{minipage}
\hfill
\begin{minipage}[t]{0.48\textwidth}
    \centering
    \includegraphics[width=\textwidth]{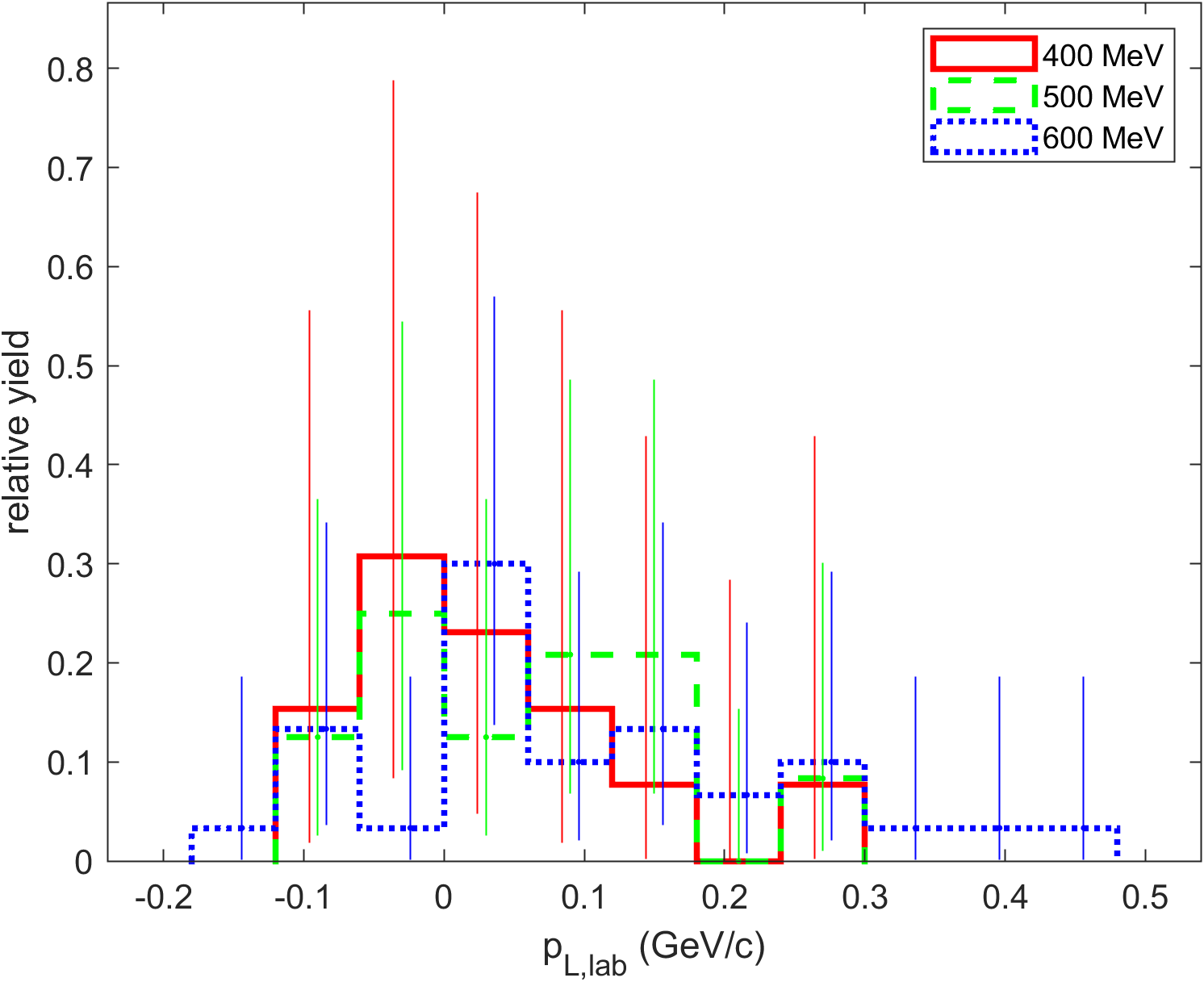}
    \caption{Normalized longitudinal momentum distributions of the produced $\pi^+$ mesons in the laboratory frame for the reaction $^{12}\mathrm{C}\,(^{12}\mathrm{C},\,^{12}\mathrm{N}\,\pi^{+})\,^{12}\mathrm{Be}$ at incident energies of $400$, $500$, and $600~\mathrm{A\,MeV}$.}
    \label{figpiplon}
\end{minipage}
\end{figure}

\begin{figure}[h]
\centering
\begin{minipage}[t]{0.48\textwidth}
    \centering
    \includegraphics[width=\textwidth]{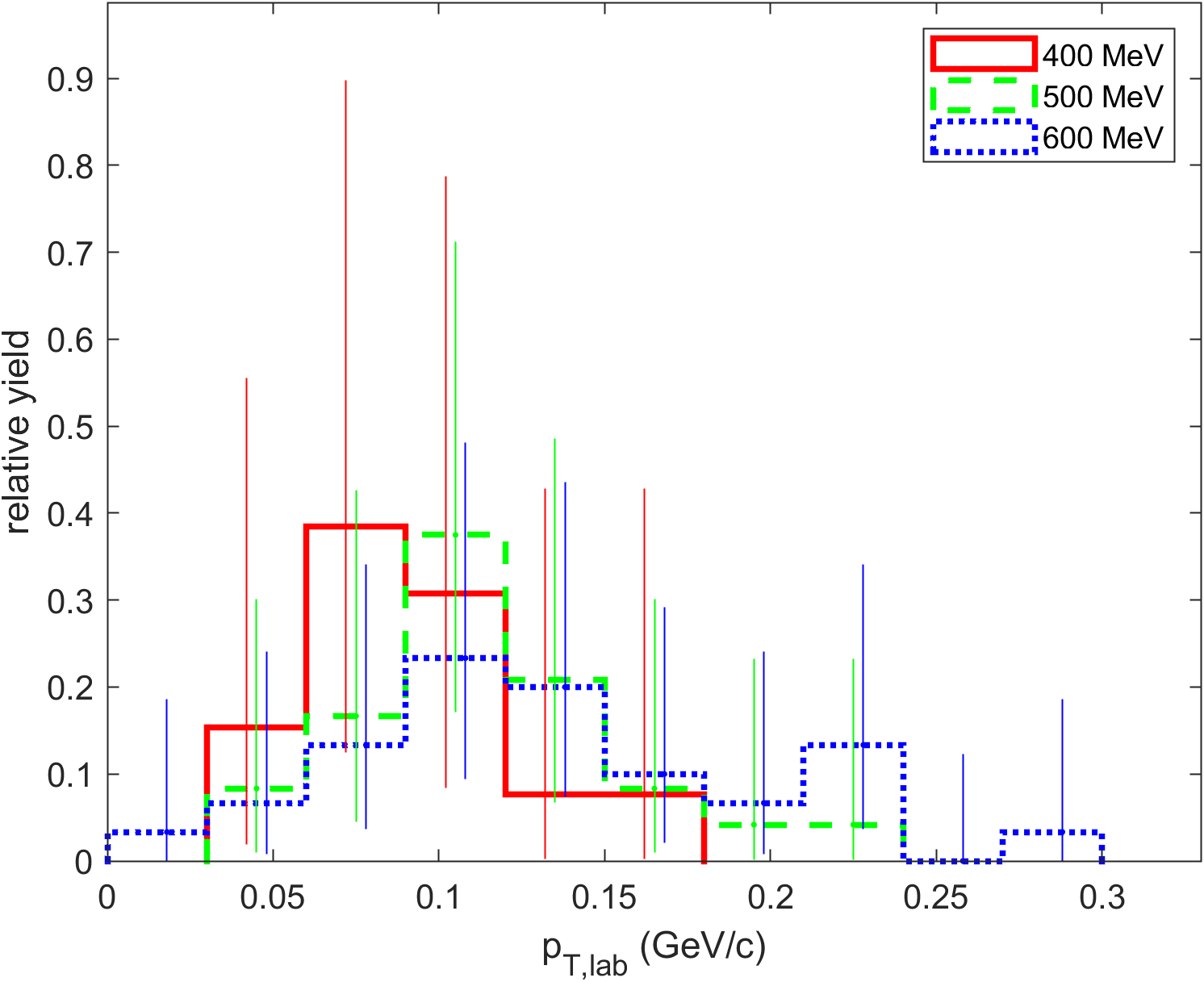}
    \caption{Normalized transverse momentum distributions of the produced $\pi^+$ mesons in the laboratory frame for the reaction $^{12}\mathrm{C}\,(^{12}\mathrm{C},\,^{12}\mathrm{N}\,\pi^{+})\,^{12}\mathrm{Be}$ at incident energies of $400$, $500$, and $600~\mathrm{A\,MeV}$.}
    \label{figpiptra}
\end{minipage}
\hfill
\begin{minipage}[t]{0.48\textwidth}
    \centering
    \includegraphics[width=\textwidth]{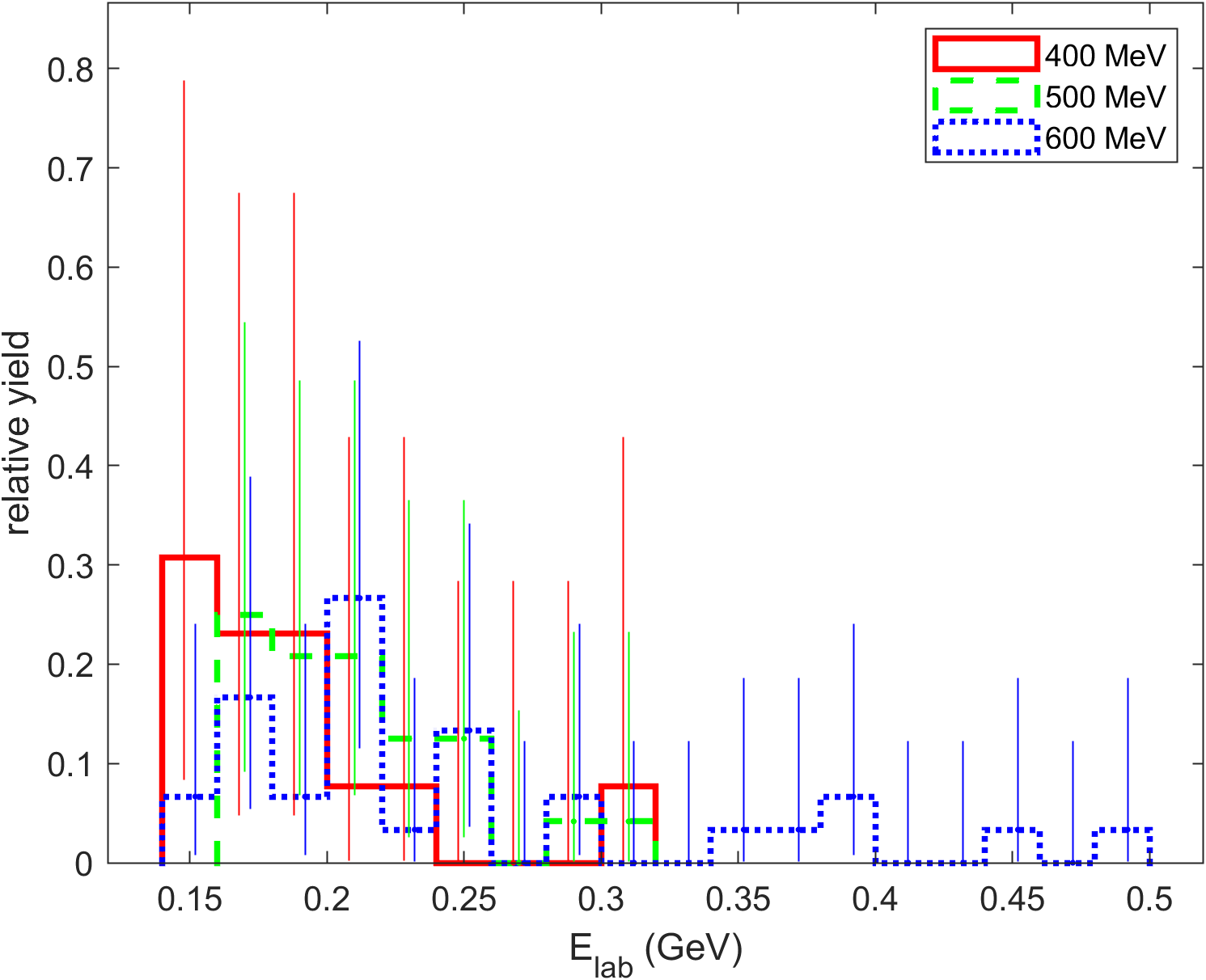}
    \caption{Normalized energy distributions of the produced $\pi^+$ mesons in the laboratory frame for the reaction $^{12}\mathrm{C}\,(^{12}\mathrm{C},\,^{12}\mathrm{N}\,\pi^{+})\,^{12}\mathrm{Be}$ at incident energies of $400$, $500$, and $600~\mathrm{A\,MeV}$.}
    \label{figpiE}
\end{minipage}
\end{figure}

As shown in Figs.~\ref{figpiptot},\ref{figpiplon}, and \ref{figpiptra}, the produced $\pi^+$ mesons possess low total, longitudinal, and transverse momenta, with minimal variations as the incident energy increases. 
The similarity between the longitudinal and transverse distributions indicates that the $\pi^+$ motion is nearly isotropic. 
The energy distributions, shown in Fig.~\ref{figpiE}, exhibit a similarly weak dependence on the incident energy, consistent with the trends observed in the momentum distributions.

These characteristics are inconsistent with direct reaction processes, which typically yield high-momentum, forward-peaked pions, and with decays of the emitted $^{12}\mathrm{N}$ nucleus, because the $^{12}\mathrm{N}$ nucleus carries a large longitudinal momentum, and its decay products would be strongly forward-peaked in the laboratory frame.  
Instead, the observed low-momentum, quasi-isotropic distributions suggest that the $\pi^+$ are predominantly produced via the decay of $\Delta$ resonances. 
This scenario naturally explains both the weak energy dependence and near-isotropy of the momentum and energy distributions.

\subsection{
Angle Distributions for the 
$^{12}\mathrm{C}(^{12}\mathrm{C},\,^{12}\mathrm{N}\,\pi^{+})\,^{12}\mathrm{Be}$ Reaction
}

The emission angle $\theta$ is defined as the angle between the momentum vector of an emitted particle and the incident beam direction.
It serves as a key observable for characterizing the spatial correlation between particle emission and beam incidence, and it provides important guidance for the optimal placement of detectors in subsequent experiments. 

In this work, the emission angles of the emitted $^{12}\mathrm{N}$ nuclei and $\pi^{+}$ mesons in the 
$^{12}\mathrm{C}(^{12}\mathrm{C},\,^{12}\mathrm{N}\,\pi^{+})\,^{12}\mathrm{Be}$ reaction are calculated as \begin{equation} \theta = \arctan \left( \frac{|\vec{p}_{T}|}{|\vec{p}_{L}|} \right), \end{equation} where $|\vec{p}{L}|$ and $|\vec{p}{T}|$ denote the longitudinal and transverse momentum components relative to the incident beam, respectively.

\begin{figure}[h]
\centering
\includegraphics[width=0.7\textwidth]{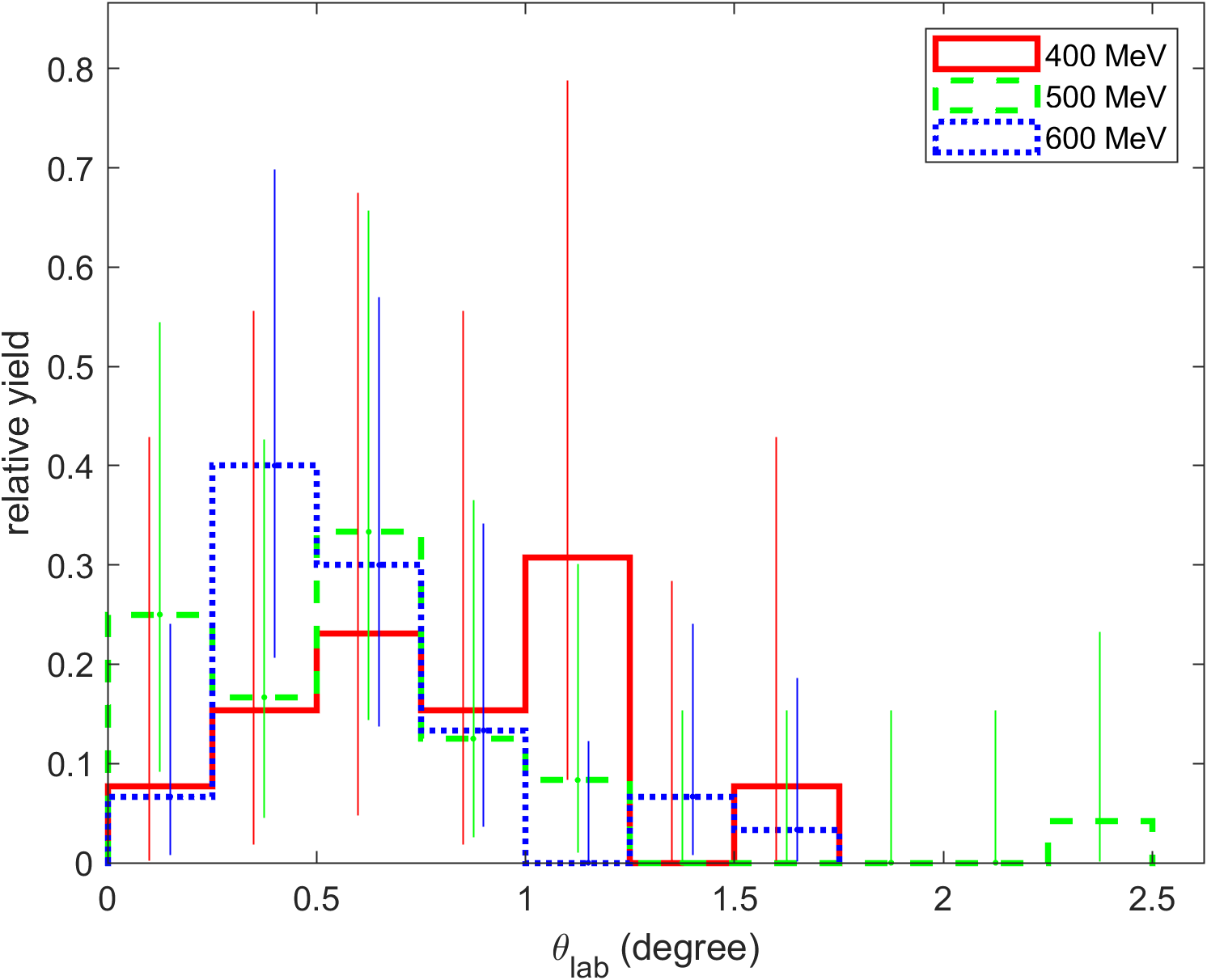} 
\caption{
Normalized angle distributions of the emitted $^{12}\mathrm{N}$ nuclei in the laboratory frame for the reaction 
$^{12}\mathrm{C}\,(^{12}\mathrm{C},\,^{12}\mathrm{N}\,\pi^{+})\,^{12}\mathrm{Be}$ at incident energies of $400$, $500$, and $600~\mathrm{A\,MeV}$.}
\label{figNang}
\end{figure}

In the laboratory frame, the emitted $^{12}\mathrm{N}$ nuclei are predominantly concentrated at small forward angles, ranging from $0^\circ$ to $2.5^\circ$ as shown in Fig.~\ref{figNang}.
This strong forward peaking indicates that the momentum transfer from the incident $^{12}\mathrm{C}$ nucleus occurs primarily along the beam direction, with negligible transverse components. 
Such a distribution is characteristic of peripheral, single CE reactions. 
Moreover, the distribution shifts toward even smaller angles as the incident energy increases, suggesting that at higher energies, the $^{12}\mathrm{N}$ emission becomes more collimated along the beam axis.
These observations are consistent with the conclusion that $^{12}\mathrm{N}$ is predominantly produced in peripheral collisions with small momentum transfers.

\begin{figure}[h]
\centering
\begin{subfigure}[b]{0.48\textwidth}
    \centering
    \begin{overpic}[width=\textwidth]{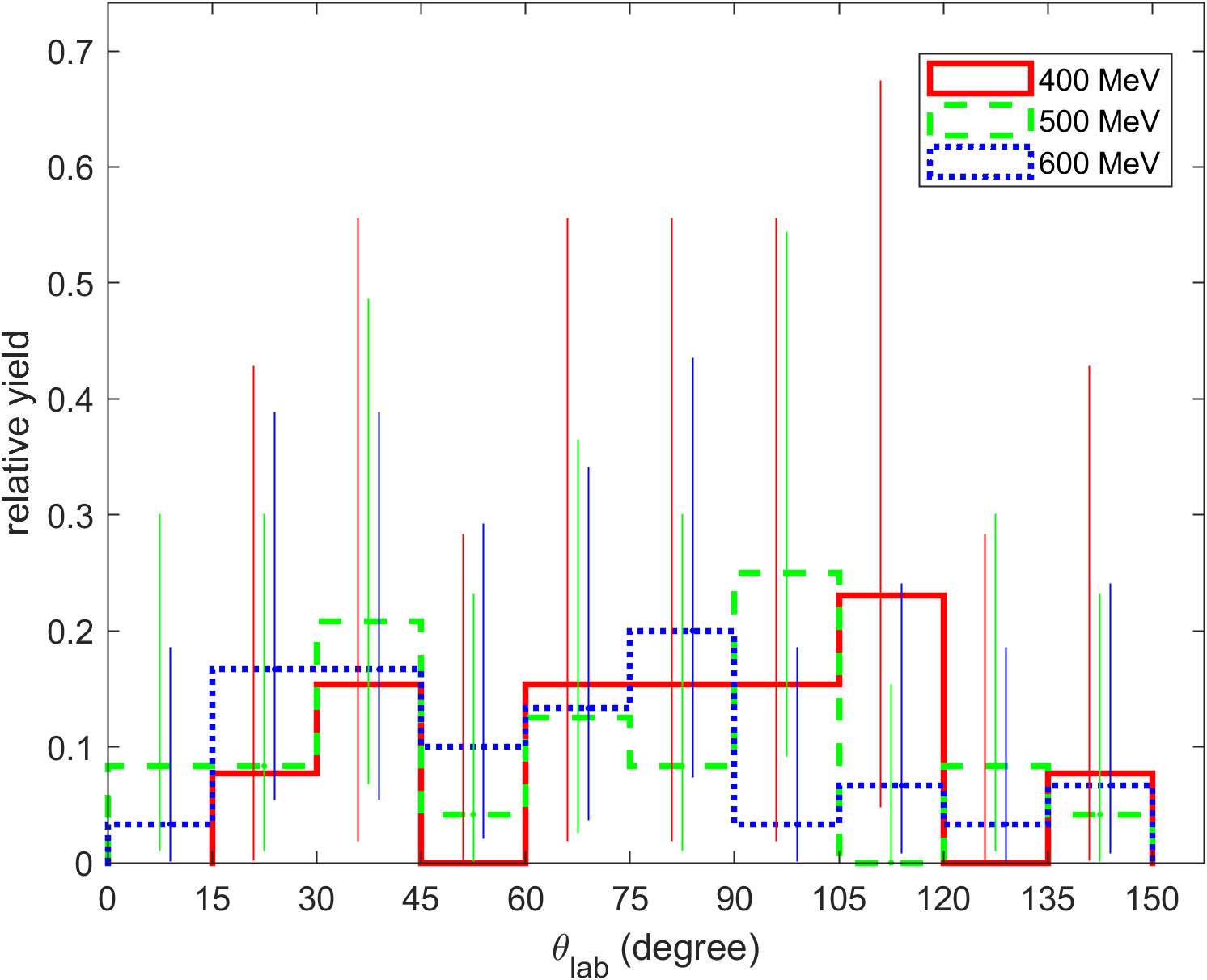}
        \put(10,74){\large\bfseries (a)} 
    \end{overpic}
    \phantomcaption 
    \label{figpiangL}
\end{subfigure}
\hfill
\begin{subfigure}[b]{0.48\textwidth}
    \centering
    \begin{overpic}[width=\textwidth]{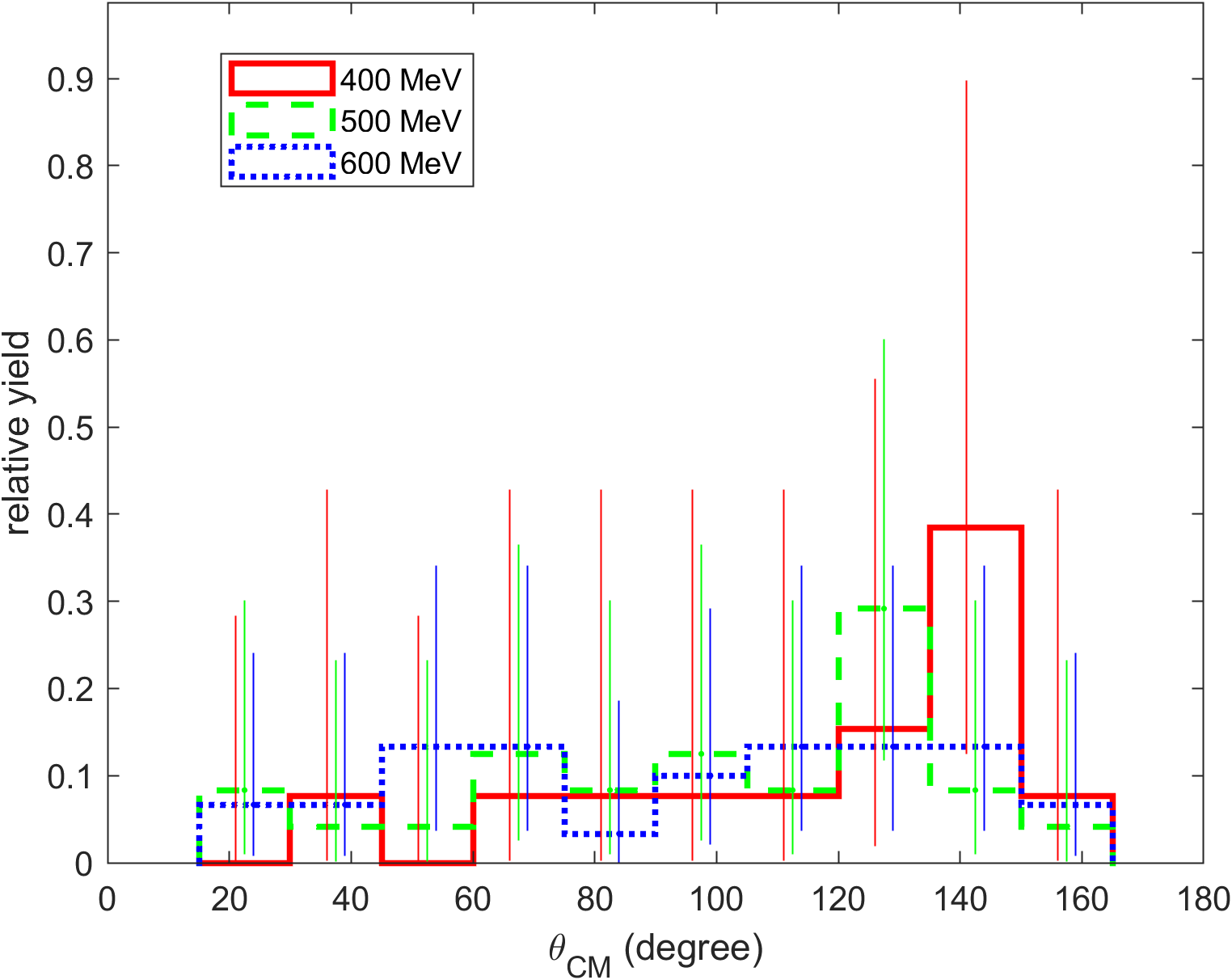}
        \put(10,74){\large\bfseries (b)} 
    \end{overpic}
    \phantomcaption
    \label{figpiangC}
\end{subfigure}

\caption{
(a) Normalized angle distributions of the produced $\pi^+$ mesons in the laboratory frame for the reaction $^{12}\mathrm{C}\,(^{12}\mathrm{C},\,^{12}\mathrm{N}\,\pi^{+})\,^{12}\mathrm{Be}$ at incident energies of $400$, $500$, and $600~\mathrm{A\,MeV}$.  
(b) Normalized angle distributions of the produced $\pi^+$ mesons in the center-of-mass frame for the reaction $^{12}\mathrm{C}\,(^{12}\mathrm{C},\,^{12}\mathrm{N}\,\pi^{+})\,^{12}\mathrm{Be}$ at incident energies of $400$, $500$, and $600~\mathrm{A\,MeV}$.}
\label{figpiang}
\end{figure}

In contrast, the emission-angle distributions of the $\pi^{+}$ mesons are much broader. 
In the laboratory frame, they extend from $0^\circ$ to approximately $150^\circ$ as shown in Fig.~\ref{figpiangL}, whereas the $^{12}\mathrm{N}$ nuclei remain strongly forward-peaked. 
If the $\pi^{+}$ mesons were produced in the same direct reaction, their angular distributions would be expected to retain a similar forward-peaked feature.
However, the observed broad distributions indicate a different production mechanism.

In the center-of-mass frame, the $\pi^{+}$ angular distributions remain broad, spanning approximately $5^\circ$ to $165^\circ$ as shown in Fig.~\ref{figpiangC}, with the majority emitted at angles greater than $90^\circ$, showing pronounced backward enhancement. 
This backward peaking weakens with increasing incident energy. Such features are consistent with a multi-step reaction scenario: the incident nucleus first interacts with target nucleons, forming an intermediate excited state, such as a $\Delta$ resonance, which then decays into a $\pi^{+}$ meson.
During this process, momentum is redistributed among nucleons, naturally resulting in the observed broad and backward-enhanced angular distributions. 
These characteristics are consistent with a multi-step reaction scenario involving the formation and decay of intermediate $\Delta$ resonances, as suggested by the momentum and energy distributions.

\subsection{
Rapidity Distributions for the 
$^{12}\mathrm{C}(^{12}\mathrm{C},\,^{12}\mathrm{N}\,\pi^{+})\,^{12}\mathrm{Be}$ Reaction
}\label{subsec2}

While the angular distributions primarily reflect the emission geometry of the reaction products, rapidity provides complementary information on their longitudinal dynamics.
Hence, we analyze the rapidity distributions of the emitted $^{12}\mathrm{N}$ nuclei and $\pi^{+}$ mesons. 
Rapidity is defined as \begin{equation} y = \frac{1}{2} \ln \frac{E + |\vec{p}_L|}{E - |\vec{p}_L|}, \end{equation} where $E$ and $p_L$ denote the total energy and longitudinal momentum of the particle, respectively.
Because rapidity differences are invariant under Lorentz boosts along the beam direction, rapidity provides a convenient and nearly frame-independent measure of longitudinal dynamics; thus, it serves as a sensitive probe of particle emission along the beam axis.

\begin{figure}[h]
\centering
\begin{subfigure}[b]{0.48\textwidth}
    \centering
    \begin{overpic}[width=\textwidth]{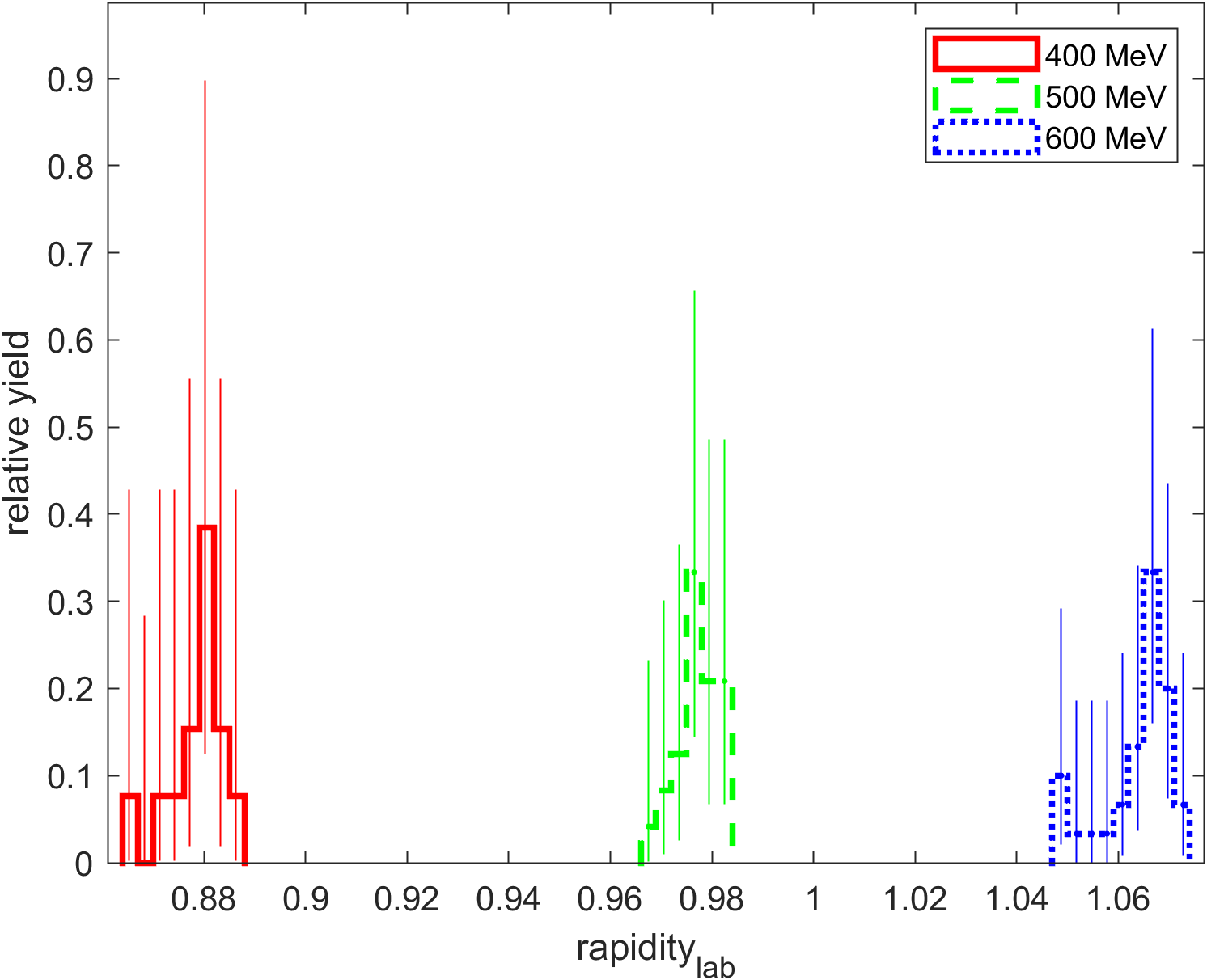}
        \put(10,74){\large\bfseries (a)} 
    \end{overpic}
    \phantomcaption 
    \label{figNrapL}
\end{subfigure}
\hfill
\begin{subfigure}[b]{0.48\textwidth}
    \centering
    \begin{overpic}[width=\textwidth]{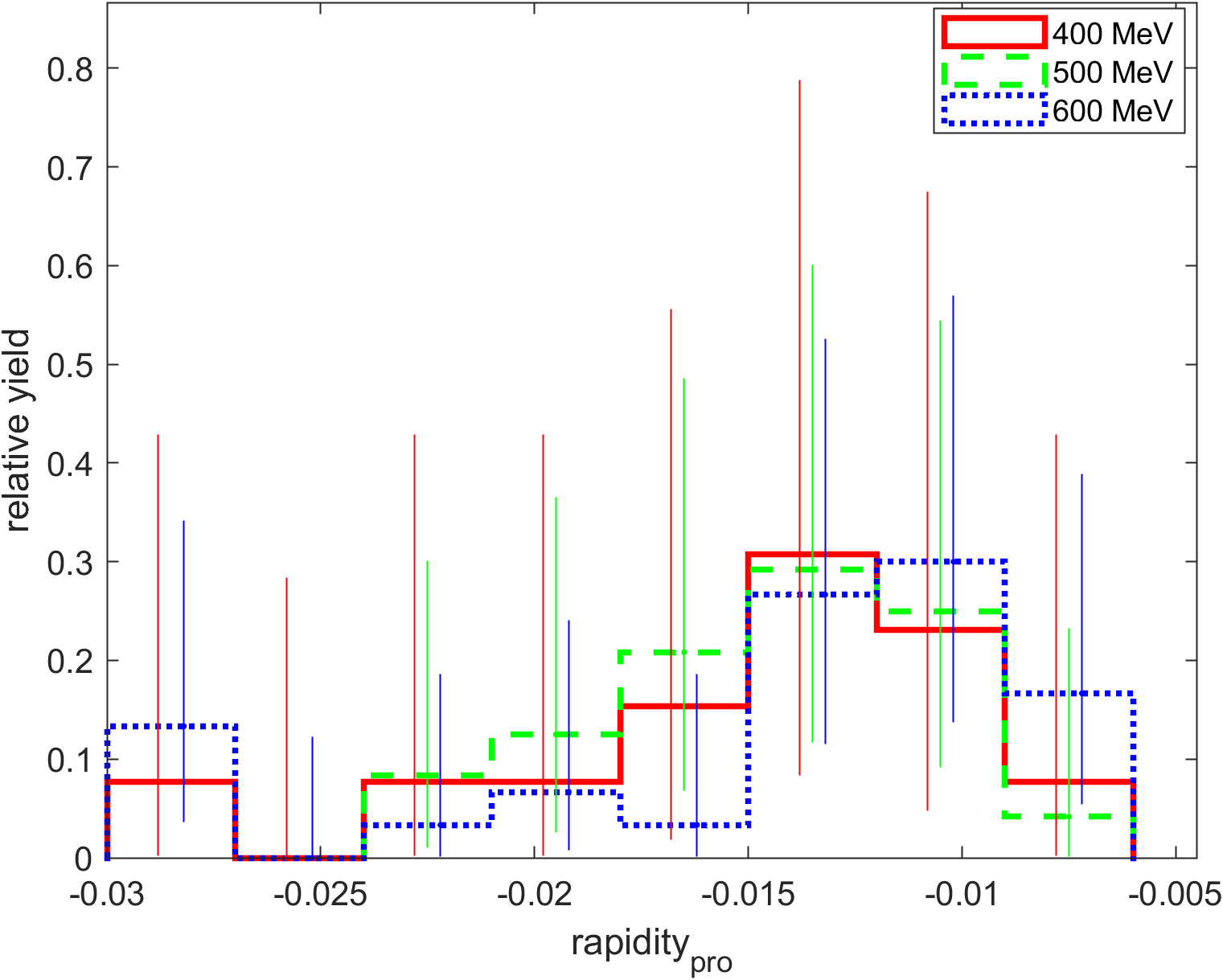}
        \put(10,74){\large\bfseries (b)} 
    \end{overpic}
    \phantomcaption
    \label{figNrapP}
\end{subfigure}

\caption{
(a) Normalized rapidity distributions of the emitted $^{12}\mathrm{N}$ nuclei in the laboratory frame for the reaction $^{12}\mathrm{C}\,(^{12}\mathrm{C},\,^{12}\mathrm{N}\,\pi^{+})\,^{12}\mathrm{Be}$ at incident energies of $400$, $500$, and $600~\mathrm{A\,MeV}$.
(b) Normalized rapidity distributions of the emitted $^{12}\mathrm{N}$ nuclei in the projectile frame for the reaction $^{12}\mathrm{C}\,(^{12}\mathrm{C},\,^{12}\mathrm{N}\,\pi^{+})\,^{12}\mathrm{Be}$ at incident energies of $400$, $500$, and $600~\mathrm{A\,MeV}$.}
\label{figNrap}
\end{figure}

From Fig.~\ref{figNrapL}, the rapidity of the emitted $^{12}\mathrm{N}$ nuclei increases with the incident beam energy. 
In contrast to central collisions, where particle production is concentrated around mid-rapidity, Fig.~\ref{figNrapP} shows that the rapidity distributions of the emitted $^{12}\mathrm{N}$ nuclei peak in the high-rapidity region, close to that of the incident nucleus. 
Moreover, the rapidities of the emitted $^{12}\mathrm{N}$ nuclei are at most only $0.03$ smaller than that of the incident $^{12}\mathrm{C}$ nucleus. 
This indicates that the formation of $^{12}\mathrm{N}$ is dominated by the projectile contribution, and the participating nucleons do not undergo significant thermalization or momentum exchange, consistent with the characteristics of peripheral collisions.

\begin{figure}[h]
\centering
\includegraphics[width=0.7\textwidth]{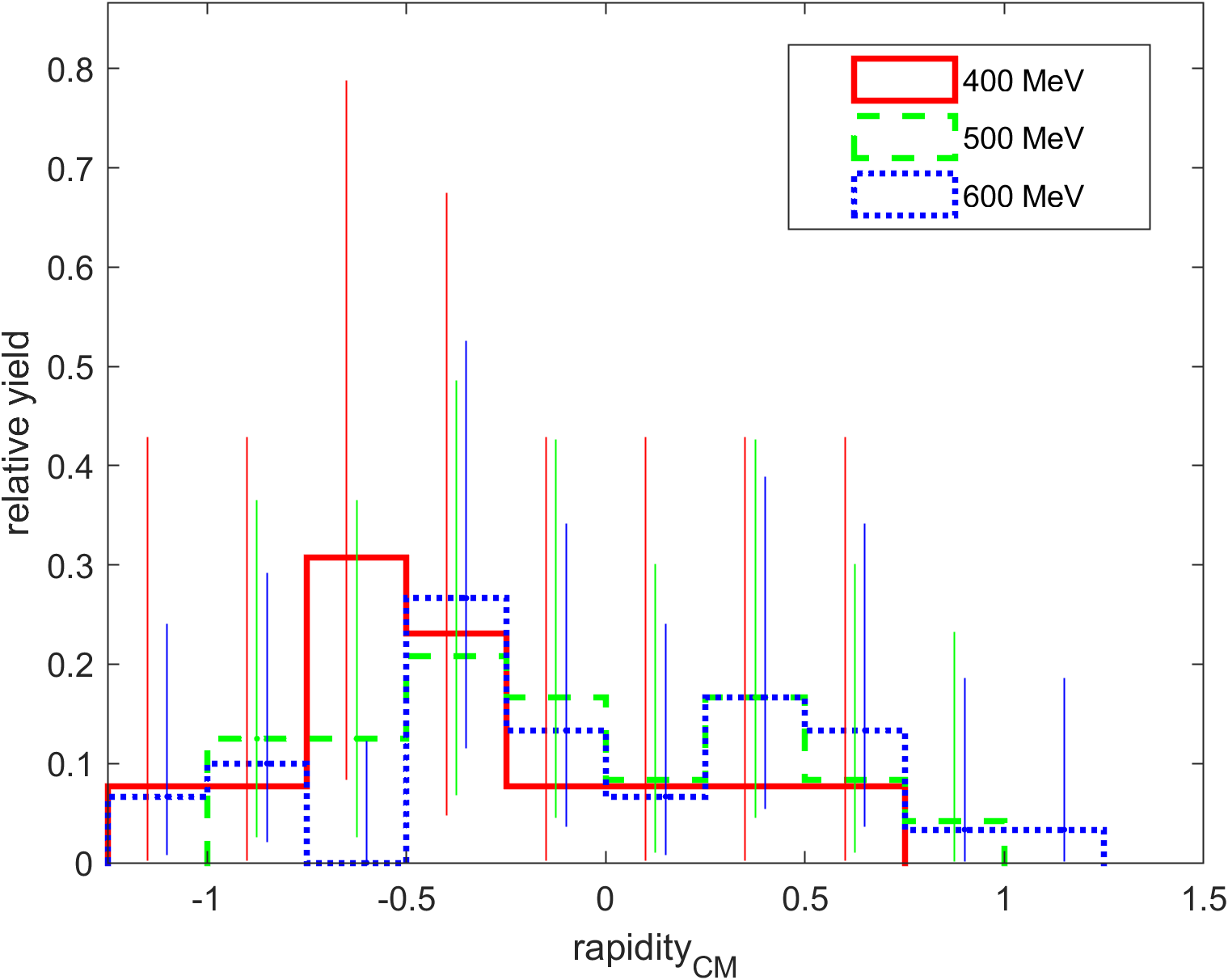} 
\caption{Normalized rapidity distributions of the produced $\pi^+$ mesons in the center-of-mass frame for the reaction $^{12}\mathrm{C}\,(^{12}\mathrm{C},\,^{12}\mathrm{N}\,\pi^{+})\,^{12}\mathrm{Be}$ at incident energies of $400$, $500$, and $600~\mathrm{A\,MeV}$.}
\label{figpirap}
\end{figure}

As shown in Fig.~\ref{figpirap}, the rapidity distributions of the $\pi^{+}$ mesons peak around $y \approx 0$, indicating that their longitudinal momenta are relatively small in the center-of-mass frame. 
A slightly higher yield is observed at negative rapidities. As the incident energy increases, the distribution shifts slightly toward positive rapidities, consistent with the weakening of the backward-emission feature observed in the angular distributions. 
The concentration around mid-rapidity suggests that these $\pi^{+}$ mesons originate predominantly from the participant region rather than from projectile-like fragments. 
This behavior is consistent with the broad and backward-enhanced angular distributions and provides further support for the interpretation, based on the momentum, energy, and angular analyses, that the $\pi^{+}$ mesons are predominantly produced via the decay of intermediate $\Delta$ resonances.

Overall, the rapidity distributions provide a complementary view of the angular emission patterns. 
The $^{12}\mathrm{N}$ nuclei exhibit rapidities close to that of the incident nucleus, reflecting their projectile-dominated origin and the peripheral character of the reaction. 
In contrast, the $\pi^{+}$ mesons are concentrated around mid-rapidity, corresponding to small longitudinal momenta and consistent with their backward-emission tendency in the center-of-mass frame. 
These features are consistent with the conclusions drawn from the momentum, energy, and angular analyses and together provide a coherent picture in which the $\pi^{+}$ mesons are predominantly produced via the decay of intermediate $\Delta$ resonances in a multi-step process.

\section{Discussion}\label{sec4}

In transport models such as UrQMD, CE-like effects may result from various microscopic mechanisms, including genuine nucleon–nucleon CE reactions and effective CE processes such as $\Delta$ resonance excitation followed by reabsorption and large-angle elastic nucleon–nucleon scattering. 
In a nuclear medium, these processes are further modified by Pauli blocking and density-dependent effects, complicating the unambiguous identification of the underlying mechanism.

The events analyzed in this paper,
\[
^{12}\mathrm{C}+^{12}\mathrm{C} \rightarrow {}^{12}\mathrm{Be} + {}^{12}\mathrm{N} + \pi^+,
\]
correspond to peripheral collisions. 
In these events, only a few nucleon–nucleon collisions occur, and the associated momentum transfer is small, making large-angle elastic scattering highly improbable. 
Although multiple $\Delta$ excitations could, in principle, contribute, their probability at the present energies is negligible. 
Moreover, additional scatterings or excitations would result in stronger nuclear excitation and reduce the probability that the residual nuclei remain bound \cite{Bondorf1995}. 
Although a complete disentanglement of all mechanisms is not possible within the UrQMD model, the combined constraints from the collision geometry, residue stability, and a single $\pi^{+}$ emission favor a scenario dominated by a nucleon–nucleon CE reaction mediated by a single $\Delta$ excitation.

Furthermore, because the cross section of the 
$^{12}\mathrm{C}(^{12}\mathrm{C},^{12}\mathrm{N}\,\pi^{+})^{12}\mathrm{Be}$
reaction is much smaller than those of other channels in $^{12}\mathrm{C}+^{12}\mathrm{C}$ collisions, a careful evaluation of the background is required to ensure that the selected events correspond to the desired reaction channel.
Therefore, the missing–mass method is employed to reconstruct the unobserved system via energy–momentum conservation:
\begin{equation}
E_{\text{miss}} = E_{\mathrm{beam}} + E_{\mathrm{tgt}} - E_{^{12}\mathrm{N}} - E_{\pi^{+}}, 
\end{equation}
\begin{equation}
\vec{P}_{\text{miss}} = \vec{P}_{\mathrm{beam}} + \vec{P}_{\mathrm{tgt}} - \vec{P}_{^{12}\mathrm{N}} - \vec{P}_{\pi^{+}},
\end{equation}
\begin{equation}
M_{\text{miss}}^{2} = E_{\text{miss}}^{2} - \bigl|\vec{P}_{\text{miss}}\bigr|^{2}. 
\end{equation}
Here, $E_{\text{miss}}$, $\vec{P}_{\text{miss}}$, and $M_{\text{miss}}$ are the energy, momentum, and invariant mass of the unobserved system, respectively. 
The missing–mass distribution enables us to determine whether the reconstructed system corresponds to the expected residual nucleus and thus serves as an effective tool for event selection in reactions with complex final states \cite{bib43}.

In practice, some emitted $\pi^{+}$ mesons may remain undetected, whereas additional $\pi^{+}$ mesons may be misidentified. 
Such effects can result in the misclassification of non-target channels as target events, thereby introducing background contributions. 
To quantify this effect, we analyze the multiplicity of $\pi^{+}$ mesons as well as the production of other mesons (such as $\pi^{0}$ and $\pi^{-}$) to identify possible background sources and estimate the maximum signal-to-background ratio.

\begin{figure}[h]
\centering
\begin{minipage}[t]{0.48\textwidth} 
    \centering
    \includegraphics[width=\textwidth]{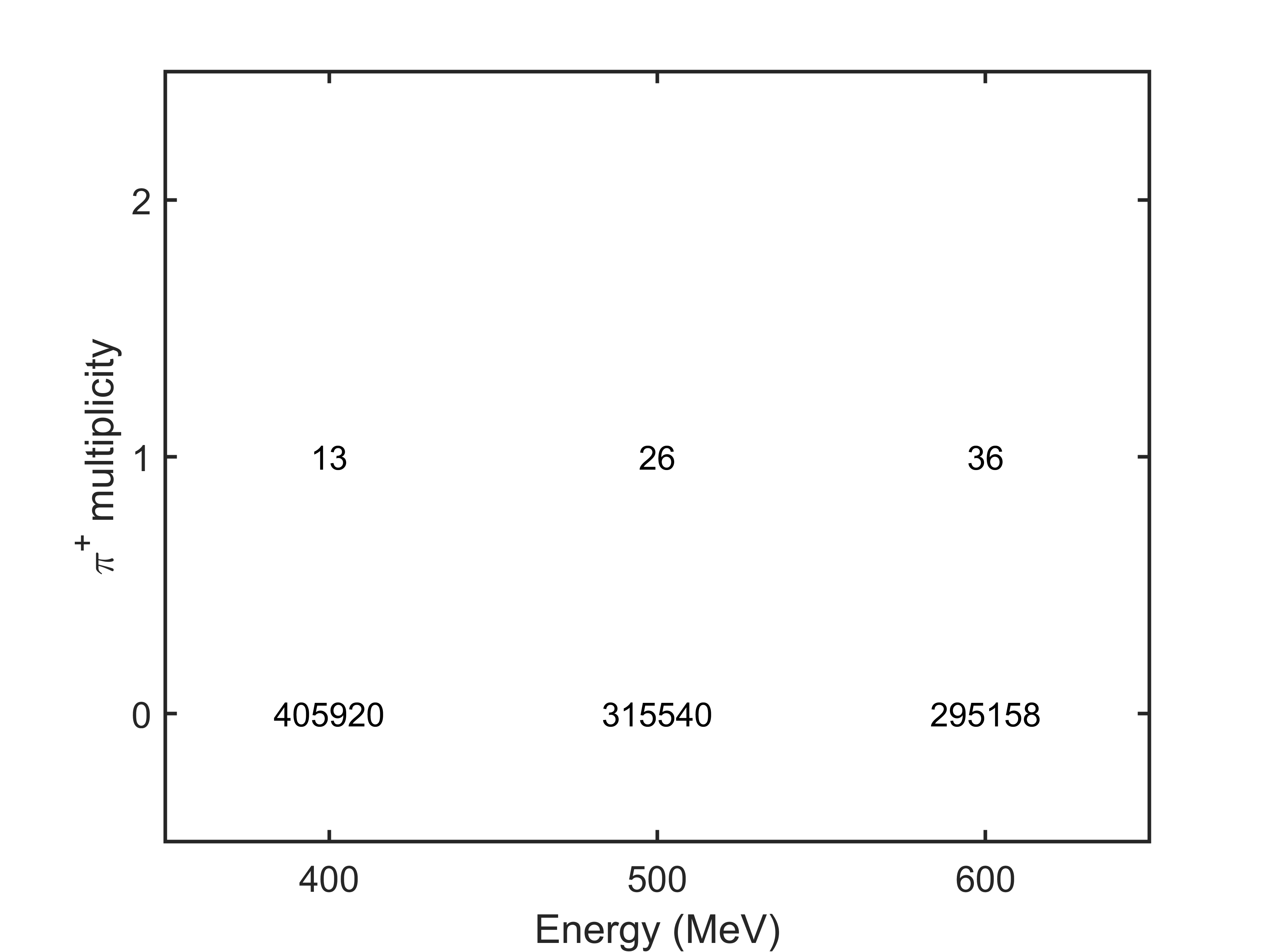}
    \caption{Multiplicity distributions of $\pi^{+}$ in the 
    $^{12}\mathrm{C}+^{12}\mathrm{C}$ 
    reaction at incident energies of 400, 500, and 600 A MeV, for events where the projectile residue is $^{12}\mathrm{N}$ and the heaviest target fragment has mass number $A=12$.}
    \label{fig10}
\end{minipage}
\hfill
\begin{minipage}[t]{0.48\textwidth} 
    \centering
    \includegraphics[width=\textwidth]{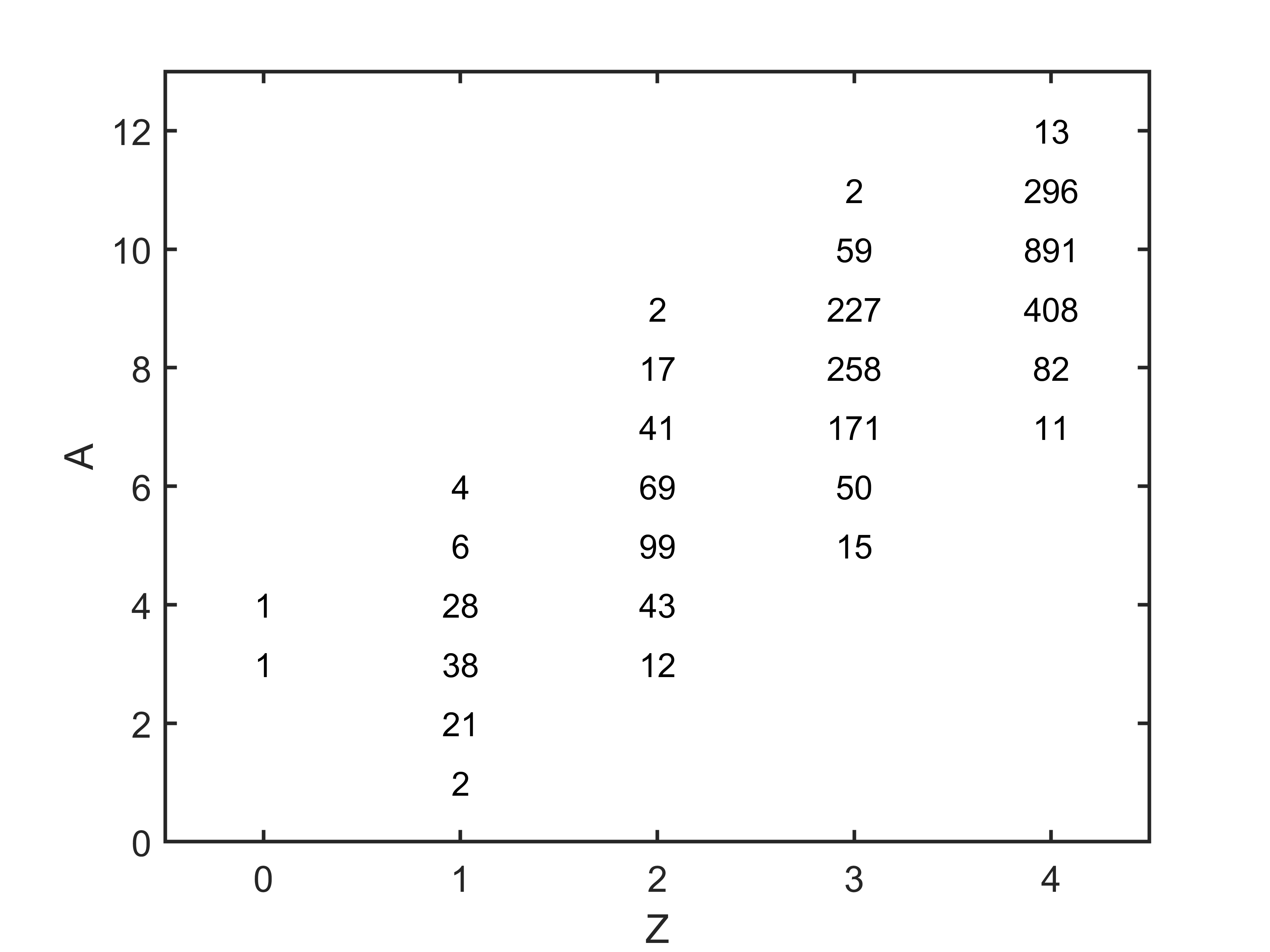}
    \caption{Correlation between the mass number $A$ and charge $Z$ of the heaviest target fragment produced in the $^{12}\mathrm{C}\,(^{12}\mathrm{C},\,^{12}\mathrm{N}\,\pi^{+})$ reaction at an incident energy of 400~A\,MeV.}
    \label{fig11}
\end{minipage}
\end{figure}

Figure~\ref{fig10} shows the multiplicity distributions of $\pi^{+}$ mesons at 400, 500, and 600~A\,MeV under the condition that the residual projectile nucleus is $^{12}\mathrm{N}$ and the heaviest target fragment has a mass number of $A=12$. 
At all three energies, no events with two $\pi^{+}$ mesons are observed. 
Furthermore, the numbers of events in which one $\pi^{+}$ meson is produced together with other mesons are 0, 2, and 6 at 400, 500, and 600~A\,MeV, respectively. 

To estimate the signal-to-background ratio, we assume that the background follows a Poisson distribution and adopt the 95\% confidence upper limit for the background event number, providing a conservative estimate of the minimum signal-to-background ratio. 
The corresponding upper limits for the background are 3.00, 6.30, and 11.84 events at 400, 500, and 600~A\,MeV, respectively \cite{bib36}. 
Accordingly, the minimum signal-to-background ratios are approximately 4.33, 3.81, and 2.53, respectively.

\begin{figure}[h]
\centering
\begin{minipage}[t]{0.48\textwidth} 
    \centering
    \includegraphics[width=\textwidth]{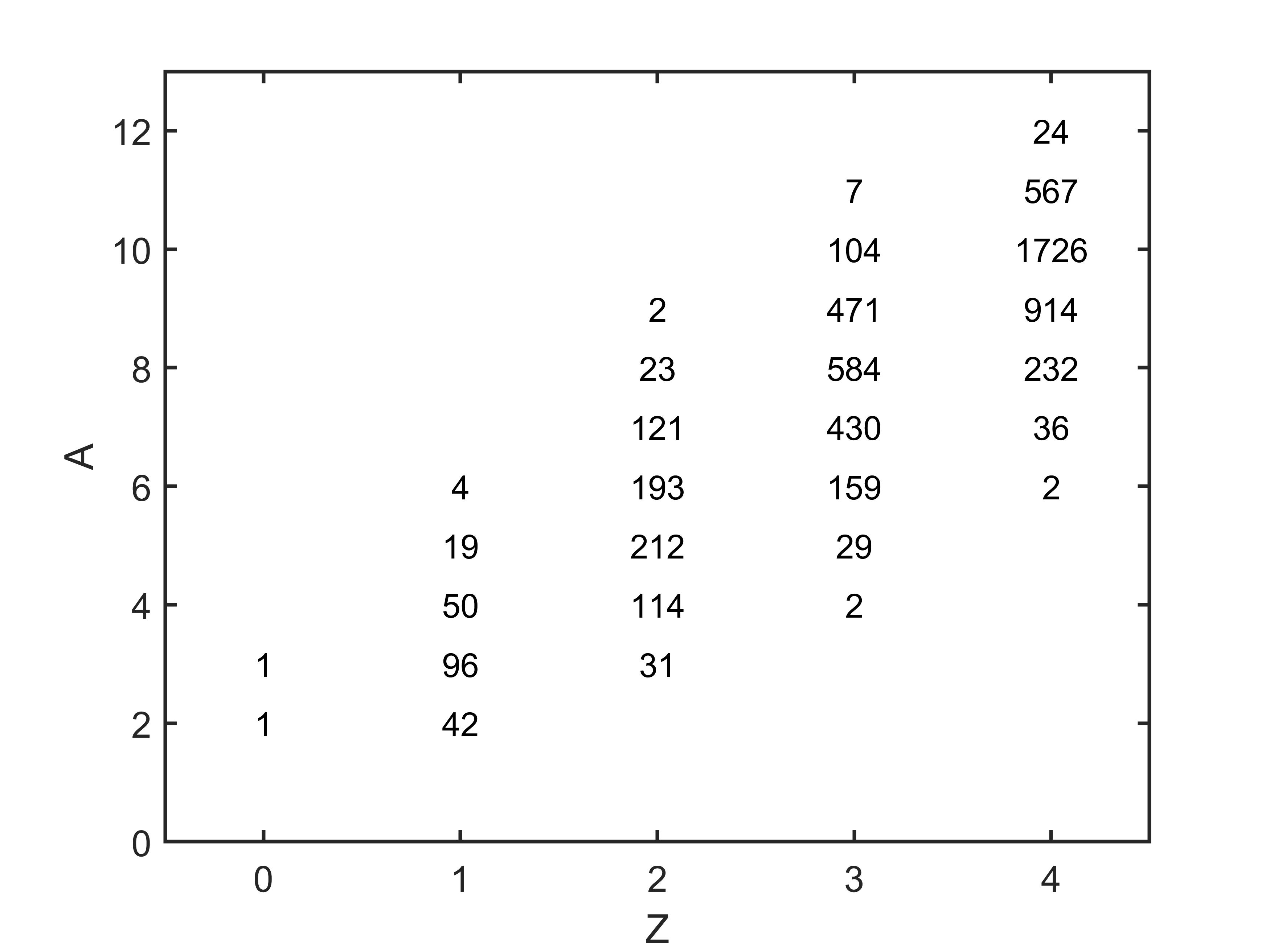}
    \caption{Correlation between $A$ and $Z$ of the heaviest target fragment produced in the $^{12}\mathrm{C}\,(^{12}\mathrm{C},\,^{12}\mathrm{N}\,\pi^{+})$ reaction at an incident energy of 500~A\,MeV.}
    \label{fig12}
\end{minipage}
\hfill
\begin{minipage}[t]{0.48\textwidth} 
    \centering
    \includegraphics[width=\textwidth]{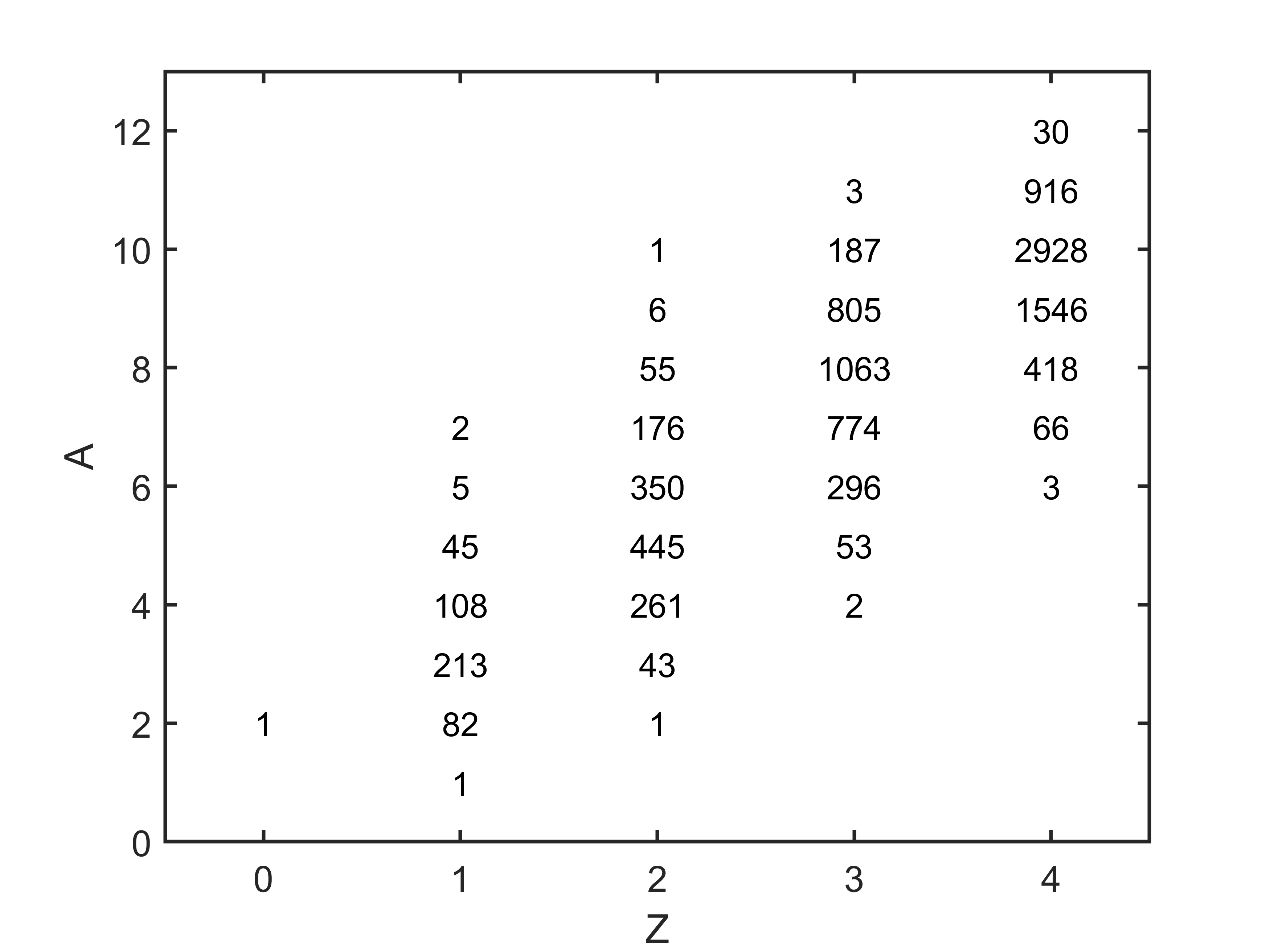}
    \caption{Correlation between $A$ and $Z$ of the heaviest target fragment produced in the $^{12}\mathrm{C}\,(^{12}\mathrm{C},\,^{12}\mathrm{N}\,\pi^{+})$ reaction at an incident energy of 600~A\,MeV.}
    \label{fig13}
\end{minipage}
\end{figure}

In addition to the background resulting from detection inefficiencies, the residual part of the target nucleus is unlikely to remain as a single $^{12}\mathrm{Be}$ nucleus; instead, it exists in multi-fragment states such as $^{11}\mathrm{Be}+n$ and $^{10}\mathrm{Be}+n+n$, as shown in Figs.~\ref{fig11}, \ref{fig12}, and \ref{fig13}. 
In missing-mass reconstruction, these channels introduce additional background, potentially affecting the purity and resolution of the spectra. 

To estimate the impact of such channels on the target reaction, we employ the UrQMD model and consider a conservative, worst-case scenario. 
Events in which the heaviest target fragment has mass number 11 are treated as
\[
^{12}\mathrm{C}+^{12}\mathrm{C} \rightarrow ^{12}\mathrm{N} + \pi^{+} + ^{11}\mathrm{Be}+n,
\] 
and events with heaviest fragment mass number 10 as
\[
^{12}\mathrm{C}+^{12}\mathrm{C} \rightarrow ^{12}\mathrm{N} + \pi^{+} + ^{10}\mathrm{Be}+n+n.
\] 

Under this assumption, at 400, 500, and 600~A\,MeV, the counting ratios of the target channel 
\[
^{12}\mathrm{C}+^{12}\mathrm{C} \rightarrow ^{12}\mathrm{N} + \pi^{+} + ^{12}\mathrm{Be}
\] 
to the channels with $^{11}\mathrm{Be}+n$ are $4.36 \times 10^{-2}$, $4.18 \times 10^{-2}$, and $3.37 \times 10^{-2}$, respectively; the corresponding ratios to the channels with $^{10}\mathrm{Be}+n+n$ are $1.37 \times 10^{-2}$, $1.31 \times 10^{-2}$, and $9.63 \times 10^{-3}$.

\section{Conclusion}\label{sec5}
In summary, we have performed a comprehensive theoretical investigation of the 
$^{12}\mathrm{C}(^{12}\mathrm{C},\,^{12}\mathrm{N}\,\pi^{+})\,^{12}\mathrm{Be}$ reaction using the UrQMD model combined with a phase-space coalescence approach. 

This work provides a novel investigation of a nontrivial heavy-ion CE channel in which CE processes and $\Delta$ resonance excitation occur simultaneously.
While these mechanisms have primarily been studied in lepton-induced or light-ion reactions, this paper extends the investigation to a many-body heavy-ion environment within the UrQMD model.
This reaction, featuring a single $\pi^+$ emission with $^{12}\mathrm{N}$ predominantly in its ground state, provides a well-defined final-state configuration suitable for studying the reaction dynamics involving both CE processes and $\Delta$ resonance excitation and decay.

Our kinematic analysis reveals that the momentum, energy, emission angle, and rapidity of $^{12}\mathrm{N}$ are strongly correlated with those of the incident $^{12}\mathrm{C}$ nucleus. This indicates that the reaction is dominated by peripheral, low-excitation collisions. 
In contrast, $\pi^+$ mesons exhibit relatively low momentum and energy, broad angular and rapidity distributions, and a backward-emission tendency in the center-of-mass frame, consistent with production via intermediate $\Delta$ resonance decay. 

Overall, this paper provides guidance for rare-isotope production and experimental design, including detector configuration, beam-energy selection, and event-identification strategies. 
Future research will incorporate experimental data to further constrain the model and improve our understanding of heavy-ion CE reactions and associated meson-production mechanisms.

\bmhead{Acknowledgements}

The authors gratefully acknowledge the support provided by the Supercomputing Center of Lanzhou University. 

\bmhead{Funded}

This work was financially supported by the National Key Research and Development Program of China (Grant No. 2022YFE0103900) and the Fundamental Research Funds for the Central Universities (Grant No. lzujbky-2023-stlt01). 

\bibliography{main}

\end{document}